\documentclass[letterpaper, 10 pt, conference]{ieeeconf}  % Comment this line out if you need a4paper
\usepackage{cite}
\usepackage{times}  % DO NOT CHANGE THIS
\usepackage{helvet} % DO NOT CHANGE THIS
\usepackage{courier}  % DO NOT CHANGE THIS
\usepackage[hyphens]{url}  % DO NOT CHANGE THIS
\usepackage{graphicx} % DO NOT CHANGE THIS
\usepackage{diagbox}
\usepackage{graphicx}
\usepackage{comment}
\usepackage{amsmath,amssymb} % define this before the line numbering.
\usepackage{color}
\usepackage{subfigure}
\usepackage{indentfirst}
\usepackage{stfloats}
\usepackage{epstopdf}
\usepackage{array}
\pdfoutput=1

\usepackage{epstopdf}

\IEEEoverridecommandlockouts                              % This command is only needed if
\title{\LARGE \bf
UMLE: Unsupervised Multi-discriminator Network for Low Light Enhancement*
}

\author{Yangyang Qu$^{1}, $ Kai Chen$^{2}$, Chao Liu$^{3}$ and Yongsheng Ou$^{3}$ % <-this % stops a space
\thanks{*This work was jointly supported by National Key Research and Development Program of China under Grant 2018AAA0103001,  National Natural Science Foundation of China (Grants No. U1613210),  Guangdong Special Support Program (2017TX04X265) and Shenzhen Fundamental Research Program (JCYJ20170413165528221)}% <-this % stops a space
\thanks{$^{1}$Yangyang Qu, $^{2}$Kai Chen, $^{3}$Chao Liu and $^{4}$Yongsheng Ou are with the Shenzhen Institutes of Advanced Technology, Chinese Academy of Sciences,Shenzhen 518055, China. Email: {\tt\small{ys.ou}@siat.ac.cn}}%
}

\begin{document}

\maketitle
\thispagestyle{empty}
\pagestyle{empty}

%%%%%%%%%%%%%%%%%%%%%%%%%%%%%%%%%%%%%%%%%%%%%%%%%%%%%%%%%%%%%%%%%%%%%%%%%%%%%%%%
\begin{abstract}

Low-light image enhancement, such as recovering color and texture details from low-light images, is a complex and vital task. For automated driving, low-light scenarios will have serious implications for vision-based applications. To address this problem, we propose a real-time unsupervised generative adversarial network (GAN) containing multiple discriminators, i.e. a  multi-scale discriminator, a texture discriminator, and a color discriminator. These distinct discriminators allow the evaluation of images from different perspectives. Further, considering that different channel features contain different information and the illumination is uneven in the image,  we propose a feature fusion attention module. This module combines channel attention with pixel attention mechanisms to extract image features. Additionally, to reduce training time, we adopt a shared encoder for the generator and the discriminator. This makes the structure of the model more compact and the training more stable.  Experiments indicate that our method is superior to the state-of-the-art methods in qualitative and quantitative evaluations, and significant improvements are achieved for both autopilot positioning and detection results.

\end{abstract}

%%%%%%%%%%%%%%%%%%%%%%%%%%%%%%%%%%%%%%%%%%%%%%%%%%%%%%%%%%%%%%%%%%%%%%%%%%%%%%%%
\section{INTRODUCTION}

%In recent years, with the improvement of camera quality, consumers' requirements of images have been increasing. Taking pictures in low-light environment  seriously influences the visual effect, and  creates a series of difficulties, such as heavy noise, and the loss of image details.  Besides, many high-level tasks are also seriously affected, such as  image instance segmentation and depth estimation.  Low-light image enhancement algorithms need to solve not only low visibility issues, but also high noise and blur.
Enhancing the low-light image is a complex task and has important applications in many fields. For autopilot, taking images in low-light environments severely affects the visual effects and causes a large number of difficulties, such as high noise and loss of image details. For simultaneous localization and mapping(SLAM), low-light images will seriously affect the localization and mapping process. In addition, many advanced tasks such as image instance segmentation and depth estimation will be severely affected. These problems will seriously affect the driving safety of automated vehicles.

\par Over the last few decades,  researchers have developed various theories to achieve the low-light enhancement, and they are categorized into three groups. The first group is based on image histogram equalization \cite{2} and correlation algorithms \cite{5, 9}. This method is useful when both the background and foreground are both too bright or too dark. However, its disadvantage is that it may increase the contrast of background noise and reduce the contrast of useful areas. The second group is based on the retinex theory \cite{3} and its variants \cite{11, 10, 7, 17}. It adaptively enhances various types of images. However, the processing results are greatly affected by the chromatic aberration variation and noise. The third group is based on convolutional neural networks and correlation algorithms \cite{24, 62,26, 27, qu}. The use of neural network techniques eases image enhancement practice, but the training requires a substantial number of dark-bright-paired images.

%\begin{figure}[t]
%\begin{center}
%\begin{tabular}{ccccc}
%\hspace{-3.5mm}
%\includegraphics[width = 0.33\linewidth]{begin1_ori.eps} &\hspace{-5mm}
%\includegraphics[width = 0.33\linewidth]{begin2_ori.eps} &\hspace{-5mm}
%\includegraphics[width = 0.33\linewidth]{begin3_ori.eps} &\hspace{-5mm}
%\\
%\hspace{-3.5mm}
%\includegraphics[width = 0.33\linewidth]{begin1_our.eps} &\hspace{-5mm}
%\includegraphics[width = 0.33\linewidth]{begin2_our.eps} &\hspace{-5mm}
%\includegraphics[width = 0.33\linewidth]{begin3_our.eps} &\hspace{-5mm}
%\\
%
%\end{tabular}
%\end{center}
%\caption{ The enhancement effect of the representative low-light images by using our proposed method UMLE. }
%\label{fig:zong}
%\end{figure}

\begin{figure}[t]
\begin{center}
\begin{tabular}{cccc}
\hspace{-3.5mm}
\includegraphics[width = 0.5\linewidth]{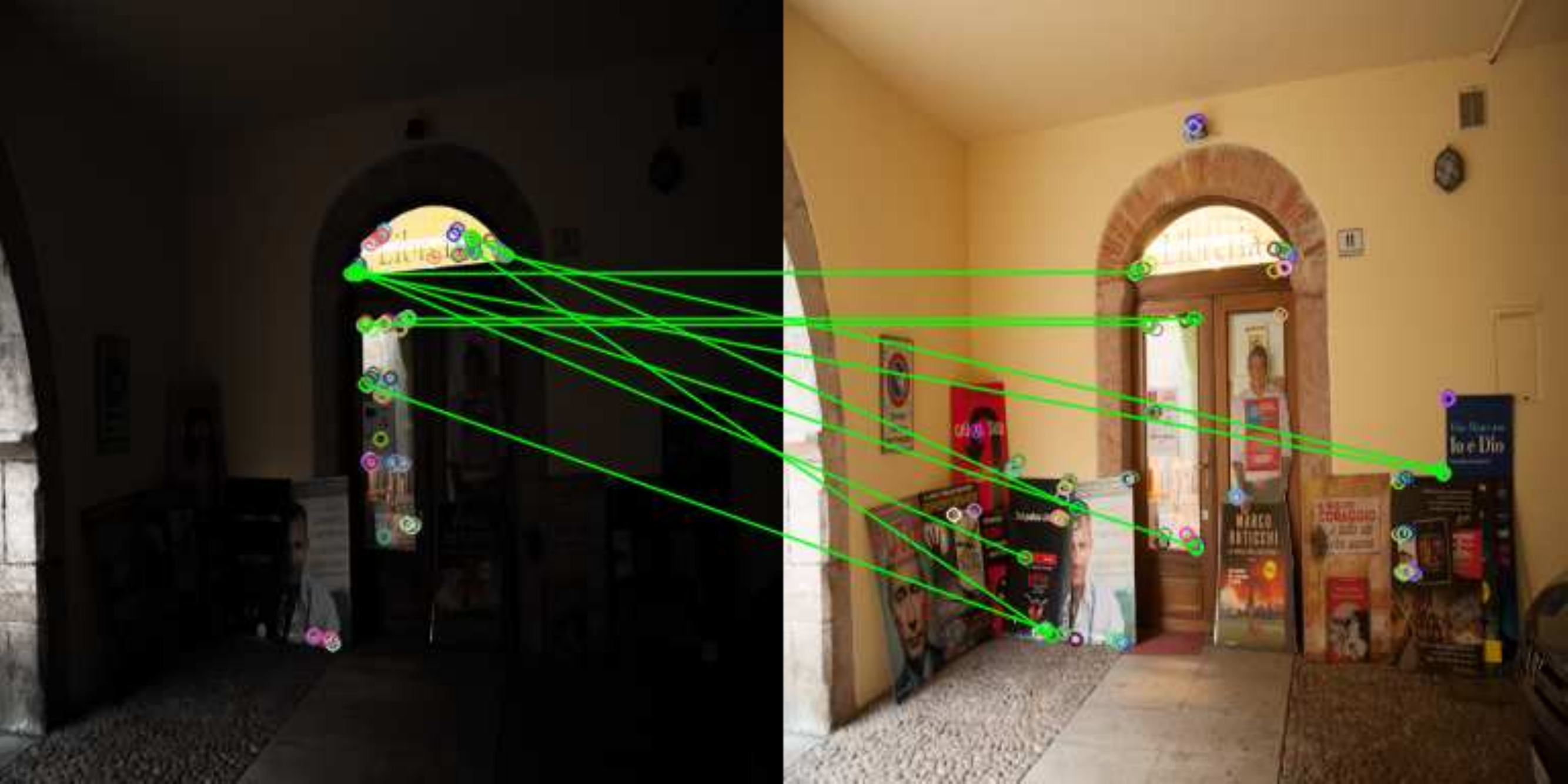} &\hspace{-5mm}
\includegraphics[width = 0.5\linewidth]{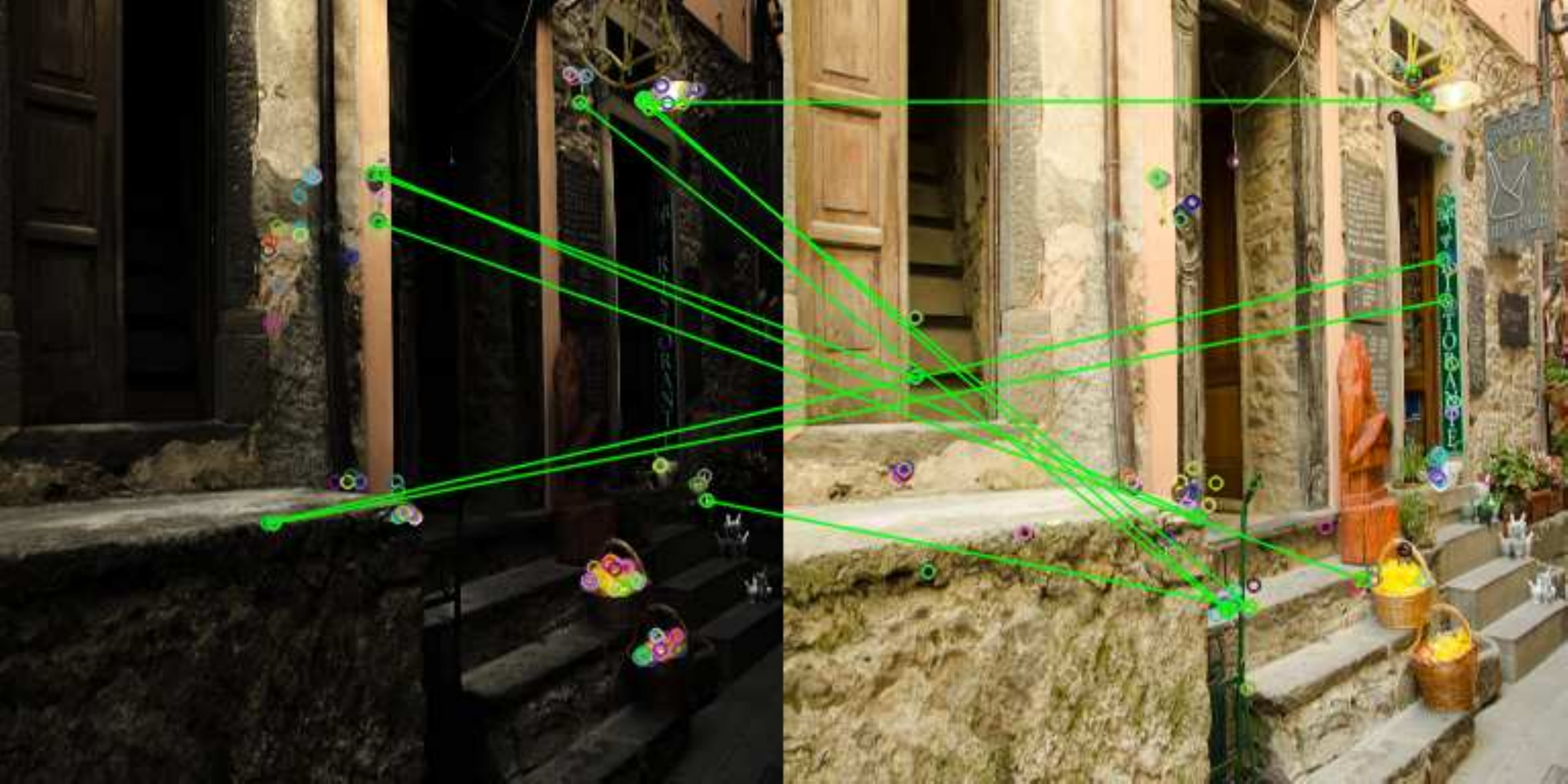} &\hspace{-5mm}
\\
\hspace{-3.5mm}
\includegraphics[width = 0.5\linewidth]{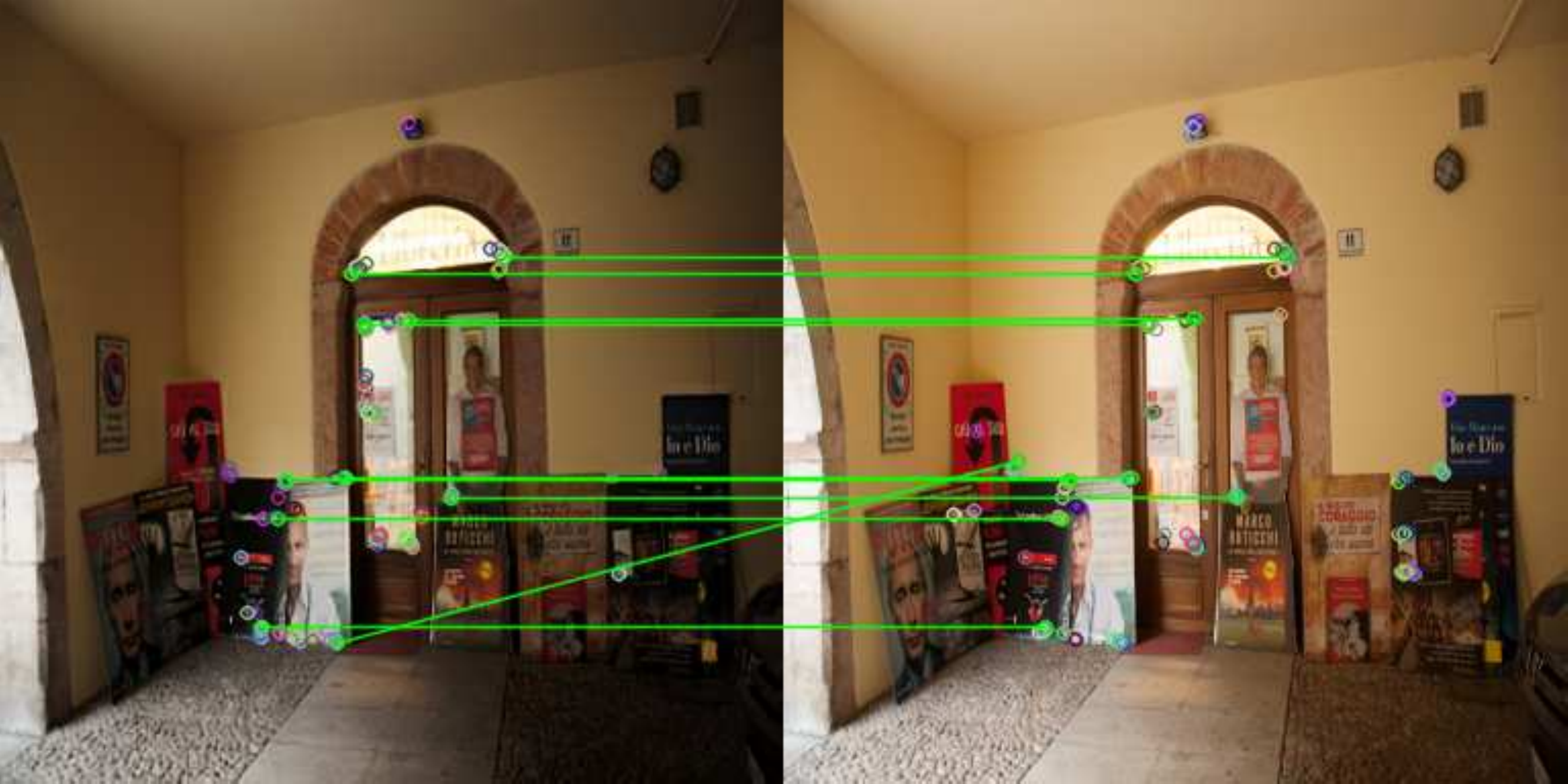} &\hspace{-5mm}
\includegraphics[width = 0.5\linewidth]{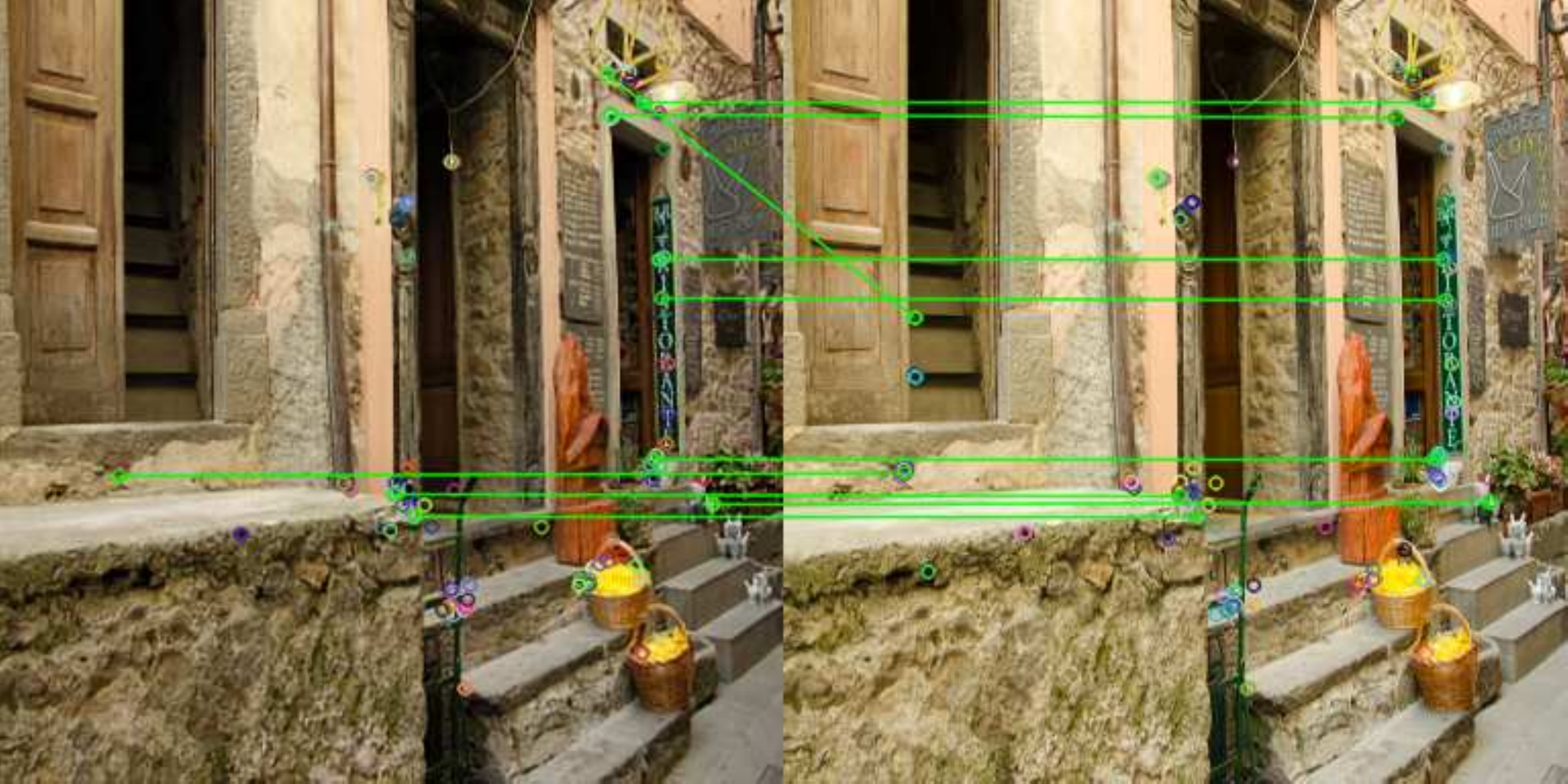} &\hspace{-5mm}
\\
(a) &\hspace{-4mm} (b)
\\
\end{tabular}
\end{center}
\caption{ Results of ORB matching after enhancement of the low-light images by our proposed real-time low-light enhancement method UMLE.}
\label{fig:zong}
\end{figure}

This paper proposes a multi-discriminator generative adversarial network (UMLE)  to restore the color and texture features of the image. The discriminator consists of three branches. The first branch is a multi-scale discriminator for which image features of different scales can be examined properly. The second branch is a color discriminator that tells whether the color in generated images look real or not. The third branch is a texture discriminator which evaluates the sharpness and clarity of edges in generated images.

We design a network that shares the encoder with the generator and the discriminator. This shared structure reduces the number of model parameters and accelerates the training.  Moreover, we design an attention module that puts emphasis on information-rich areas. Via an extensive experimental study, the results indicate our method is qualitatively and quantitatively superior to state-of-the-art methods.

The contributions of this paper are summarized as follows:
\begin{itemize}
\item We propose a low-light enhancement GAN. The model's training is independent of paired  training data, so it can use images of different scenes and different illuminations.  This can effectively improve the generalization  and  inhibit overfitting.
\item We design a multi-branch discriminator which can comprehensively evaluate the image  from color, texture, and global information. In addition, we present a novel attention module which combines the channel attention  and the pixel attention, and it helps to focus more on the low-light areas.
\item Our proposed method achieves a significant improvement in automatic driving tasks such as SLAM repositioning and driveable area detection under severe light changes.
\end{itemize}

\section{Proposed Model}
\subsection{Overview}
\begin{figure}[t]
\centering
\includegraphics[width=1\columnwidth]{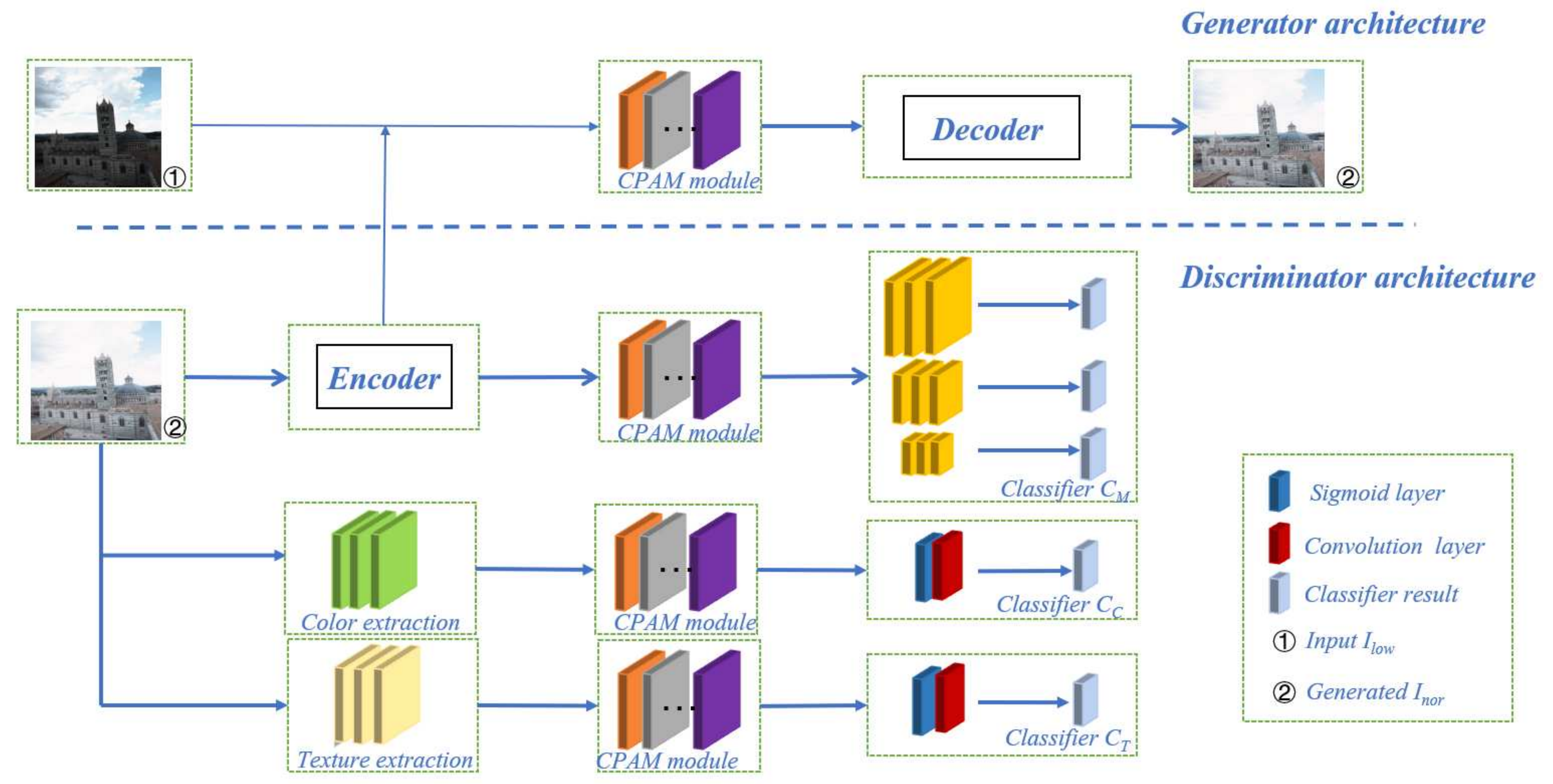} % Reduce the figure size so that it is slightly narrower than the column. Don't use precise values for figure width.This setup will avoid overfull boxes.
\caption{The architecture of our network. The proposed network consists of two branches. One branch processes the original image, and the second enhances the original image after edge detection. Specific details are described in Sec.Proposed method.}
\label{fig1}
\end{figure}

As shown in Fig.~\ref{fig1}, we design an unsupervised GAN structure which does not require paired training data to enhance the images.
Let $ L$ and $N$ be the low-light domain and the normal-light domain. $x \in$ $X_{L}, X_{N}$ indicates $x$ from the low-light domain and normal-light domain. Our model studies a transformation from the $ L$ domain to the $N$ domain.

Our model contains two generators ($G_{L\rightarrow N}$ transforms low-light image to  normal light image, and $G_{N\rightarrow L}$ transforms normal-light image to low-light image) and two discriminators ($D_{N}$ distinguishes the generated low-light images from the real low-light images, $D_{L}$ distinguishes the generated normal light images from the real normal light images.)
A generator typically includes an encoder and a decoder, whereas a discriminator includes an encoder and a classifier.  Different from other related approaches, our generator and discriminator share the same encoder. This approach has been presented in \cite{NICE}. Due to the shared structure, the training's stability is improved and the model size is reduced.

Because the images' details are largely hidden in the dark, there are color deviations, texture errors, and other problems when generating images. This paper proposes a multi-branch discriminator to solve it, which includes a color discriminator, a texture discriminator, and a multi-scaled discriminator.
\subsection{Channel and Pixel Attention Module (CPA)}
\begin{figure}[t]
\centering
\includegraphics[width=0.5\textwidth]{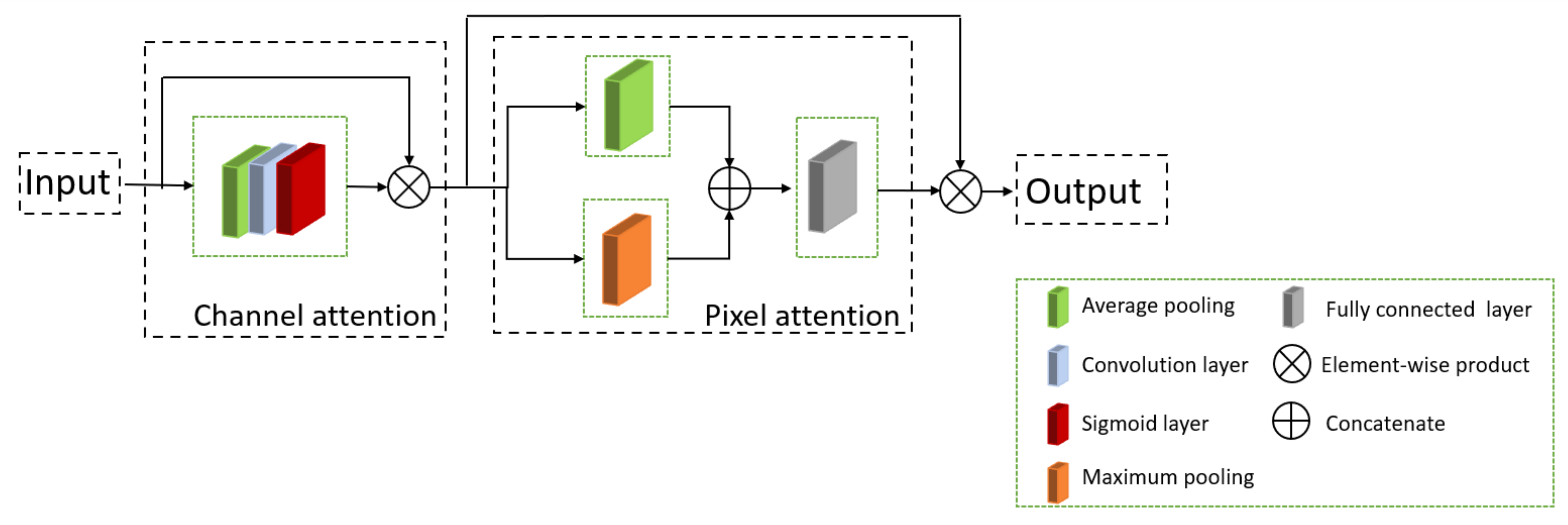} % Reduce the figure size so that it is slightly narrower than the column.
\caption{The architecture of the CPA module. It contains a channel attention module and an attention module. Specific details are described in Sec. Channel and Pixel Attention Module.}
\label{fig2}
\end{figure}

A  previous study \cite{SENet} has proved that channel attention can effectively improve the performance of  a CNN's performance. Wang et al. \cite{ECA-Net} demonstrate the avoidance of channel dimensionality can enhance the effect of channel attention. As shown in Fig.~\ref{fig2}, we present an attention module which combines the channel attention  and the pixel attention.
\par The channel attention can achieve channel attention without a dimensionality reduction. First, the module changes the input feature graph from $C\times H\times W$ to $C\times1\times 1$ through a channel-wise global average pooling, where $C, W, H$ mean channel dimension, width, and height. Suppose that $\mathcal{X} \in \mathbb{R}^{H \times W\times C}$,  the global average pooling  can be obtained by:
\begin{equation}
g_{c}(\mathcal{X})=\frac{1}{H \times W} \sum_{i=1}^{H} \sum_{j=1}^{W} \mathcal{X}_{i j}.
\end{equation}
Then, the weights of the different channels are obtained by:
\begin{equation}
\omega_{c}=\sigma\left(\operatorname{conv}\left(g_{c})\right)\right),
\end{equation}
where $\sigma$ is a sigmoid function, and $conv$ is a convolution function. Finally, the module multiplies the input by the $\omega_{c}$ to obtain the channel attention result:
\begin{equation}
c_{r}= \omega_{c} \otimes \mathcal{X}.
\end{equation}
The  pixel attention module can output the pixel attention adaptively. We splice the global average pooling and global max pooling together to obtain the pixel attention feature:
\begin{equation}
p_{r} =cat(\sigma(c_{r}),\gamma(c_{r})),
\end{equation}
where $\gamma$ is the fully connected function.
Then, the pixel attention result is:
\begin{equation}
cp_r =\gamma (p_r) \otimes c_{r},
\end{equation}
and the output is the result of the CPA module.
\par By adding a residual module in conjunction with CPA, we propose a feature extraction module (CPAM) which can dynamically output the features, and can get better obtain the pixel and channel distribution.
\subsection{Multi-branch Discriminator}
The discriminator consists of two parts, encoder $E_{D}$ and classifier $C_{D}$. Through a series of experiments, it can be found that for low-light enhancement, color information and edge information are usually not evident.  Using a single discriminator often leads to chromatic aberration, edge blur, and rising overexposure and underexposure problems in local varying illumination. To counter these problems, we put forward a  multi-branch discriminator $D_{M}$, which consists of three parts.

\subsubsection{Color discriminator} The first part is a  color discriminator $D_{c}$, and its principle design objective is to discriminate the image using the color characteristics. It is supposed to learn the contrast and color differences between the original image $I_{ori}$ and the generated image $I_{gen} $, while avoiding texture comparisons. Inspired by [6], we adopt the idea of frequency separation. For this part, we apply a  Gaussian low-pass filter:
\begin{equation}
G_{x, y}=\lambda \exp \left(-\frac{\left(x-\mu_{x}\right)^{2}}{2 \sigma_{x}}-\frac{\left(y-\mu_{y}\right)^{2}}{2 \sigma_{y}}\right),
\end{equation}
where $\lambda=0.053, \mu_{x, y}=0, \text { and } \sigma_{x, y}=3$.
Then  features are extracted through the encoder module and the CPAM module to restrain the color distortion, and dynamically output the  color features.

\subsubsection{Texture discriminator} The second part is a texture feature discriminator $D_{T}$,  and its  principle design objective is to classify the texture characteristics of the image. Previous attempts have proposed a texture classifier which emulates the process of extracting SIFT descriptors \cite{30}. They convolve it with the two filters to obtain the same $x,y$ gradients for the discriminator in a differentiable manner. In contrast, our method uses a Gaussian high-pass filter. Through this filter, we can extract the image's texture information and avoid the influence of color and other information. Then the feature is extracted through the encode module and the CPAF module, which can make our model better distinguish the image from the perspective of texture.
\begin{figure}[t]
\begin{center}
\begin{tabular}{ccccc}
\includegraphics[width = 0.235\linewidth]{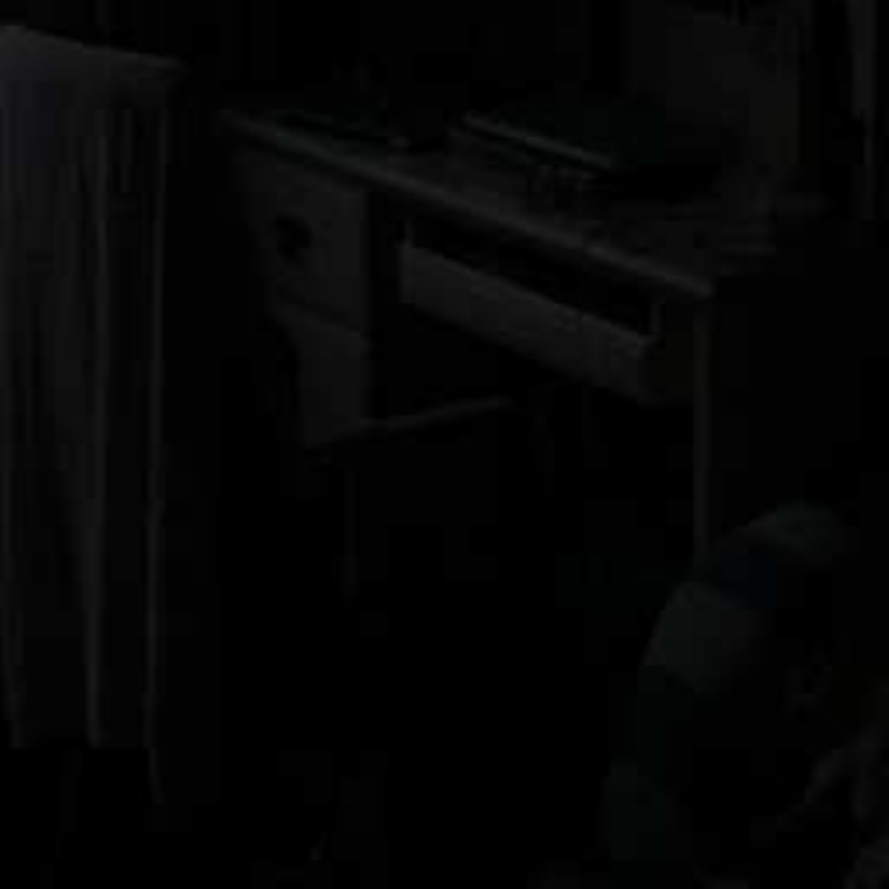} &\hspace{-4.5mm}
\includegraphics[width = 0.235\linewidth]{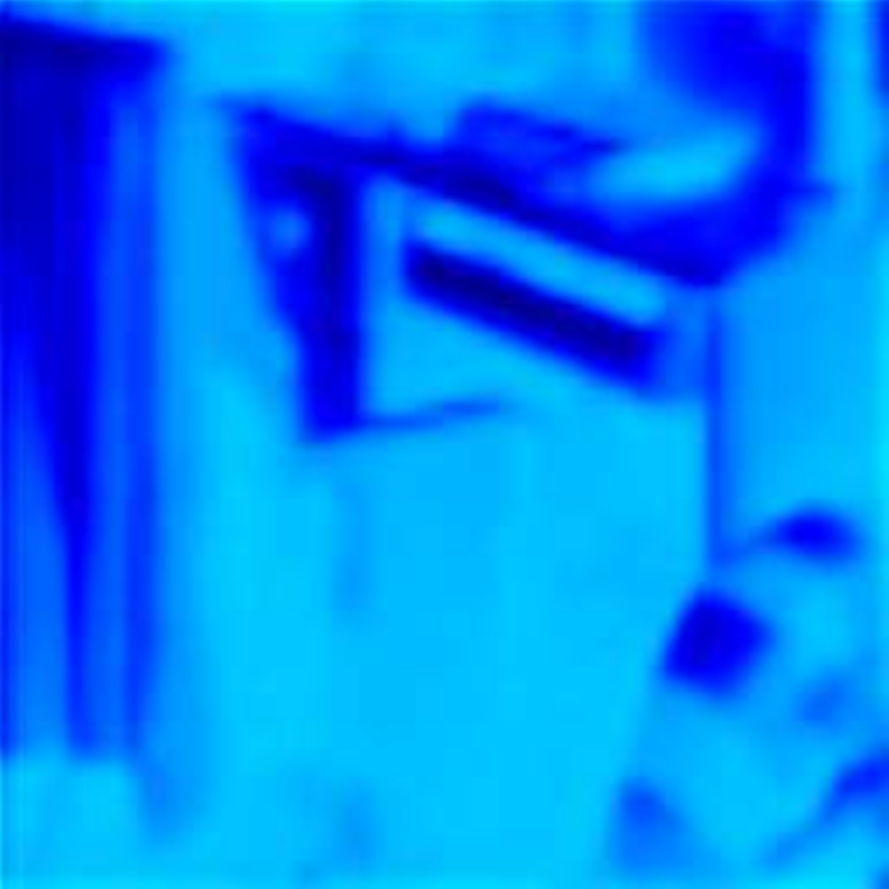} &\hspace{-4.5mm}
\includegraphics[width = 0.235\linewidth]{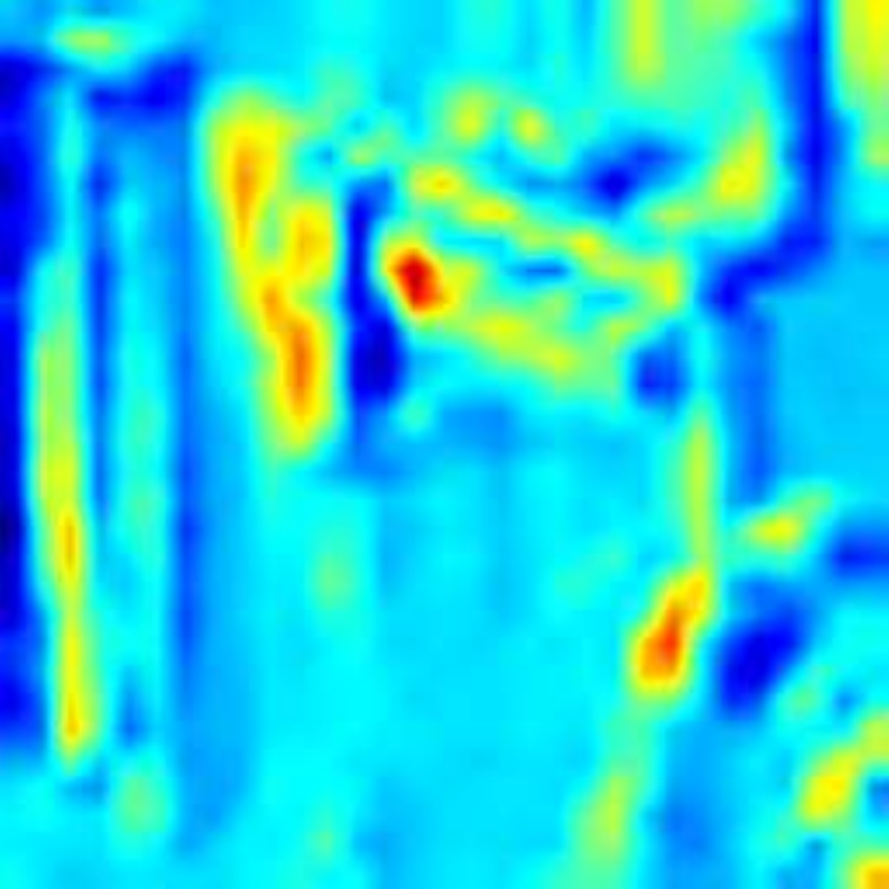}&\hspace{-4.5mm}
\includegraphics[width = 0.235\linewidth]{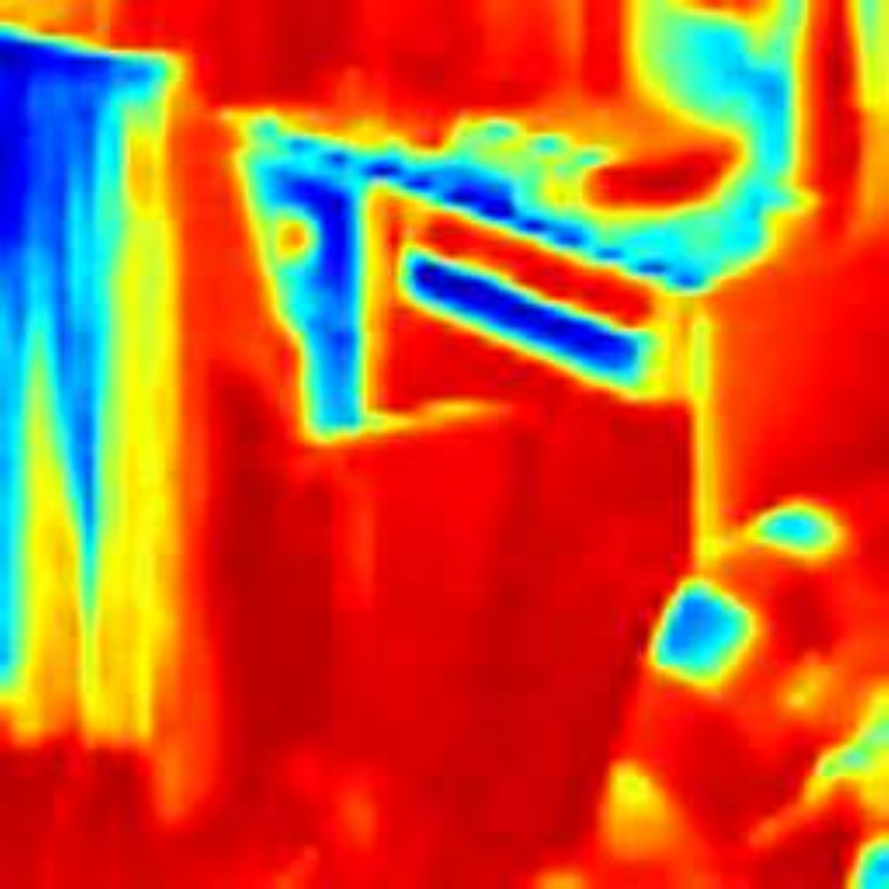}&\hspace{-4.5mm}
\\
(a) &\hspace{-4mm} (b)  &\hspace{-4mm}  (c) &\hspace{-4mm} (d)
\\
\end{tabular}
\end{center}
\caption{The feature extraction results of the three discriminators. (a) is the original image, (b) is the visualization of the color discriminator, (c) is the visualization of the texture discriminator, (d) is the visualization of the multi-scale discriminator. It can be found that the results of feature extraction of different discriminators are quite different to combine different discriminators to get better results.}
\label{fig:visualization}
\end{figure}
\subsubsection{Multi-scale discriminator} For the third part, we adopt a multi-scale discriminator $D_{S}$. Previous experiments have established that a multi-scale classifier is helpful to improve the effect of discriminators. Different from EnlightenGAN \cite{31} which uses a global-local discriminator, our model trains a three-scale classifier to judge whether the image is real or not. Details in different scales are often need to be enhanced, such as the lightened area. A global classifier cannot pay attention to these regions, for example, to enhance some details present in different scales, such as the lightened area. Providing another two scales in the classifier can focus on different sizes of regions, making the resulting image more realistic. The proposed discriminator contains a local discriminator $D_{SL}$ which is a $10*10$ pixel image patch.
\par In order to intuitively illustrate the function of the three discriminators, we visualize their results  after the CPAM module in Fig.~\ref{fig:visualization}. It can be observed that different discriminators pay different attention to the image.  The  texture  discriminator  pays  more  attention  to  the edge area and texture details of the image, while the color discriminator pays more attention to the whole image. The multi-scale discriminator can extract the color and texture features of the image at the same time.
\begin{figure*}[t]
\begin{center}
\begin{tabular}{ccccccccc}
\hspace{-4mm}h
\includegraphics[width = 0.125\linewidth]{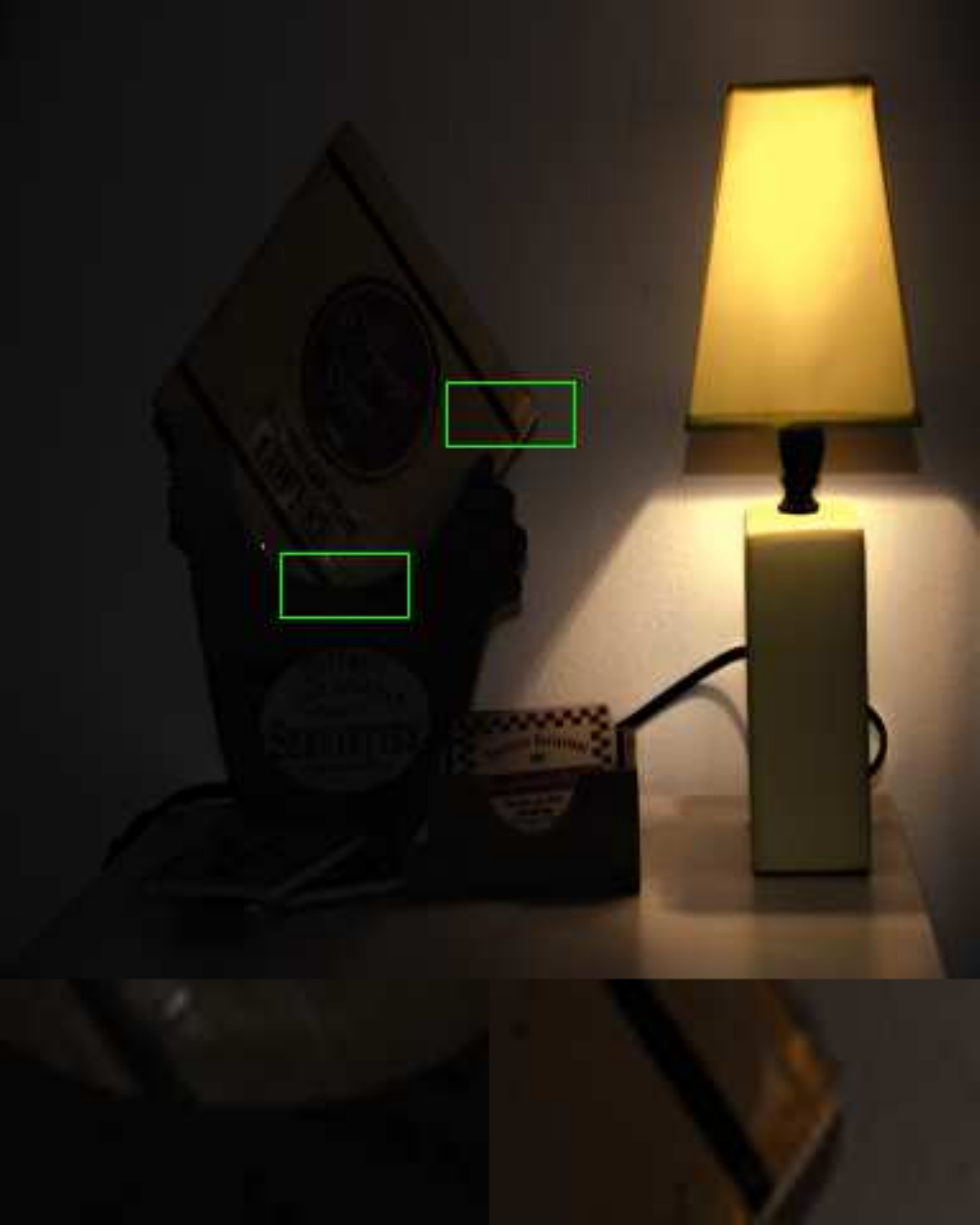} &\hspace{-5mm}
\includegraphics[width = 0.125\linewidth]{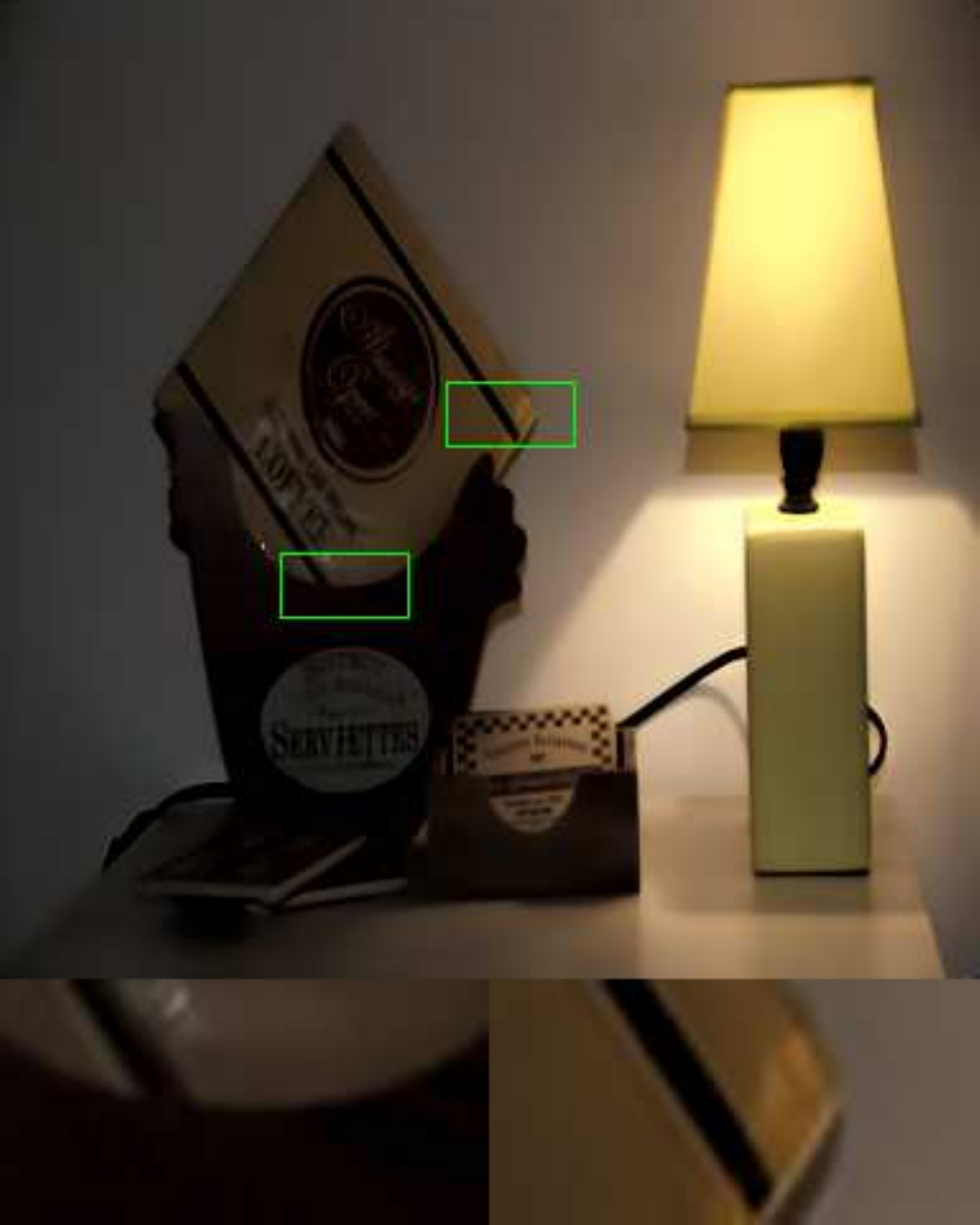}&\hspace{-5mm}
\includegraphics[width = 0.125\linewidth]{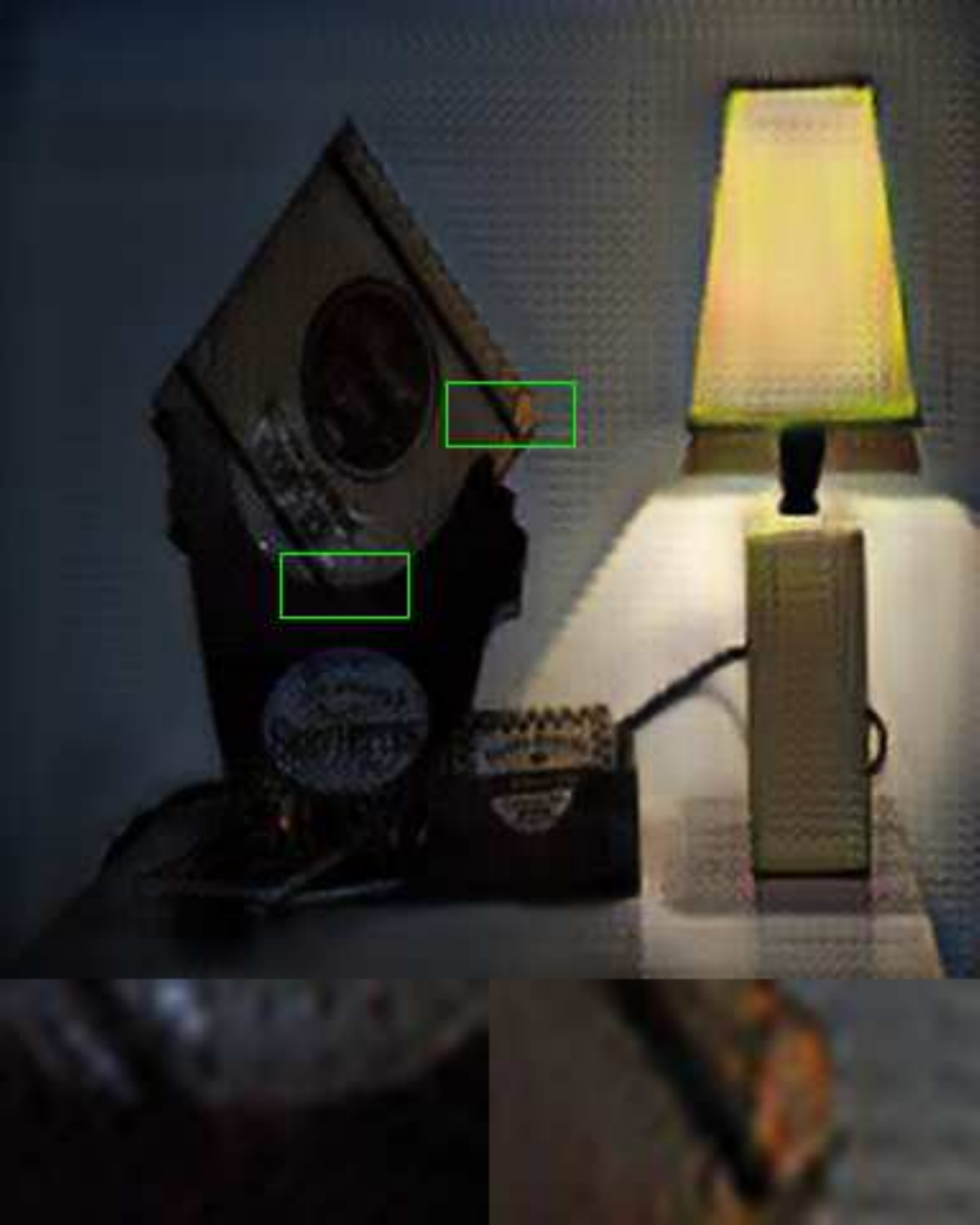}&\hspace{-5mm}
\includegraphics[width = 0.125\linewidth]{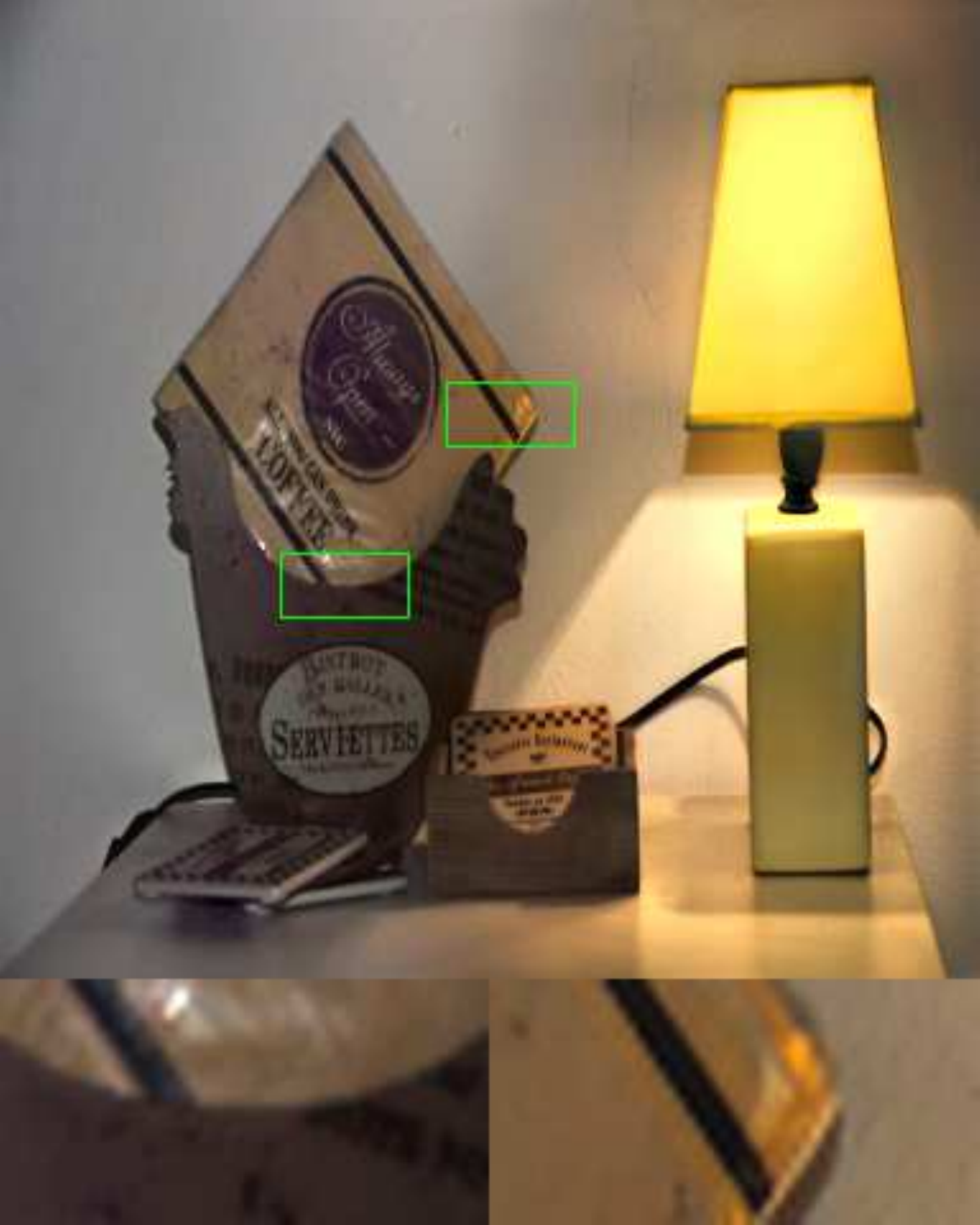}&\hspace{-5mm}
\includegraphics[width = 0.125\linewidth]{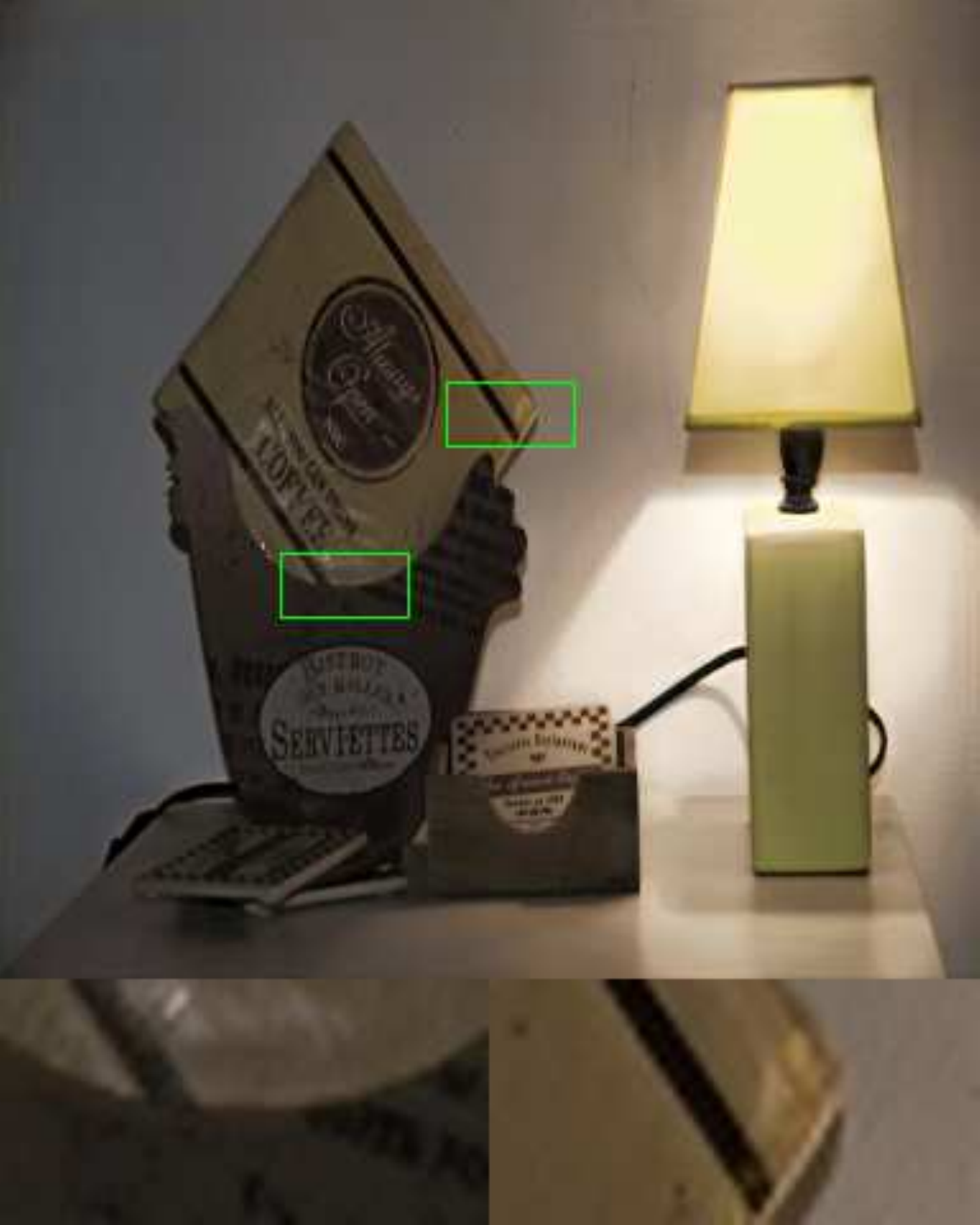}&\hspace{-5mm}
\includegraphics[width = 0.125\linewidth]{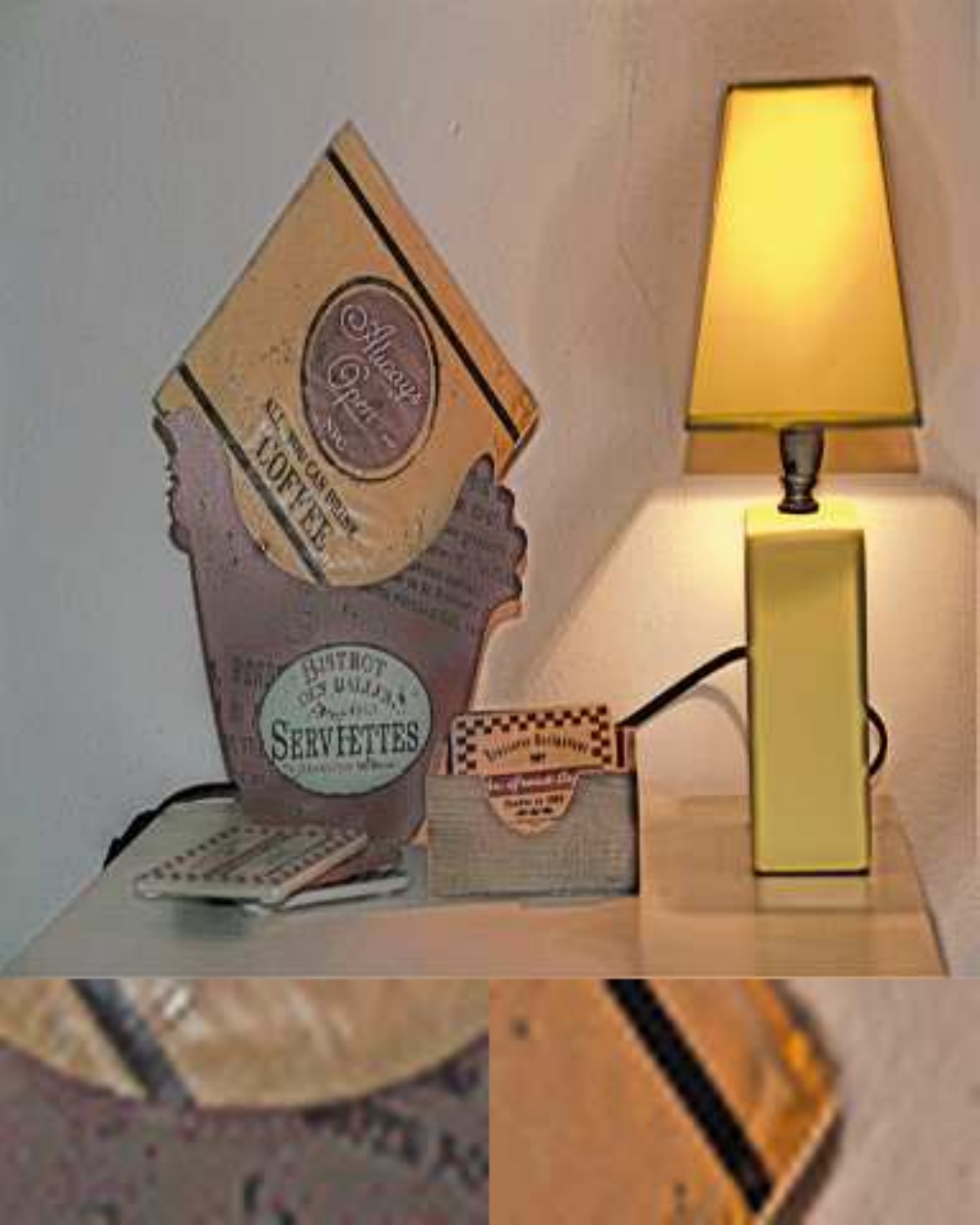}&\hspace{-5mm}
\includegraphics[width = 0.125\linewidth]{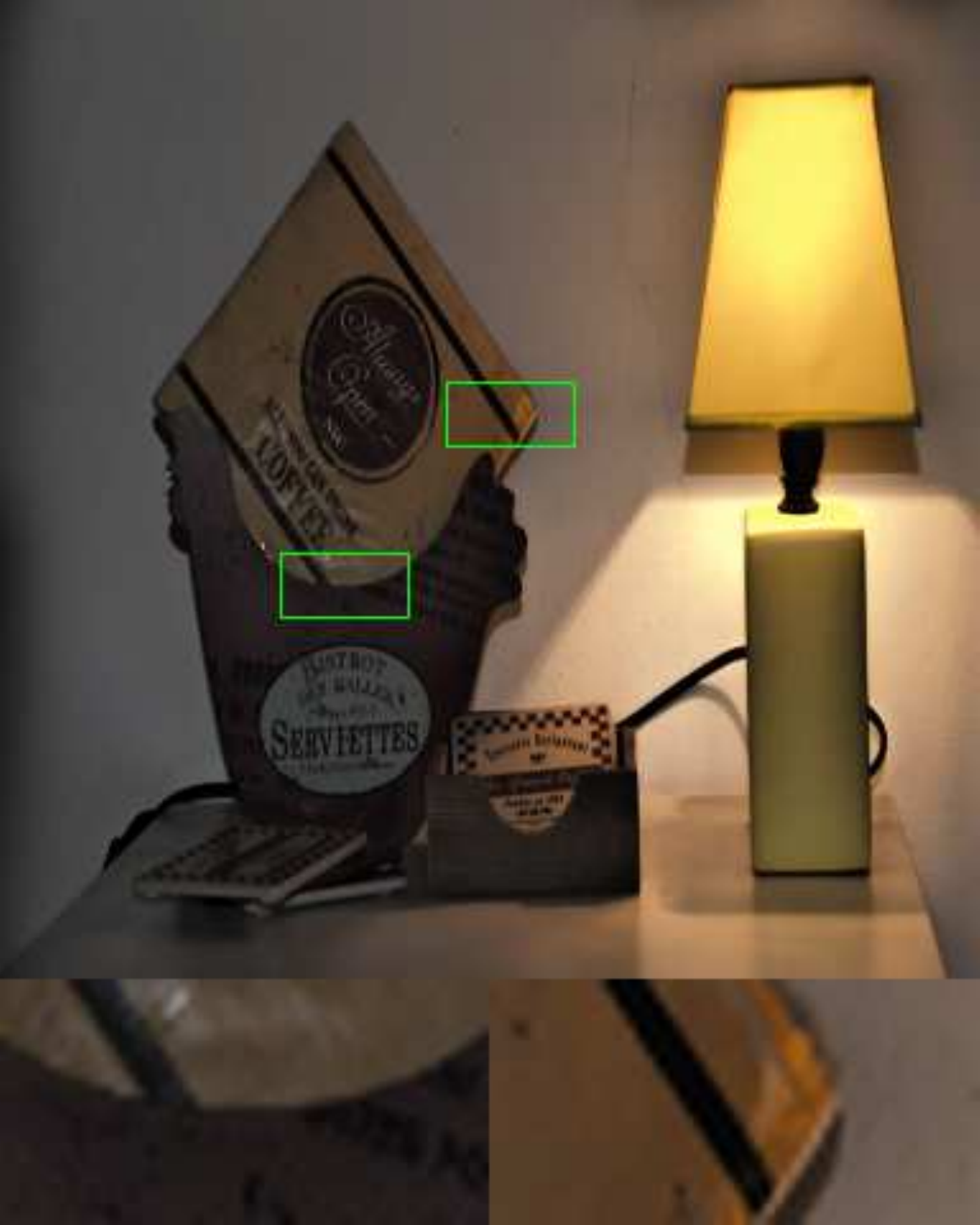}&\hspace{-5mm}
\includegraphics[width = 0.125\linewidth]{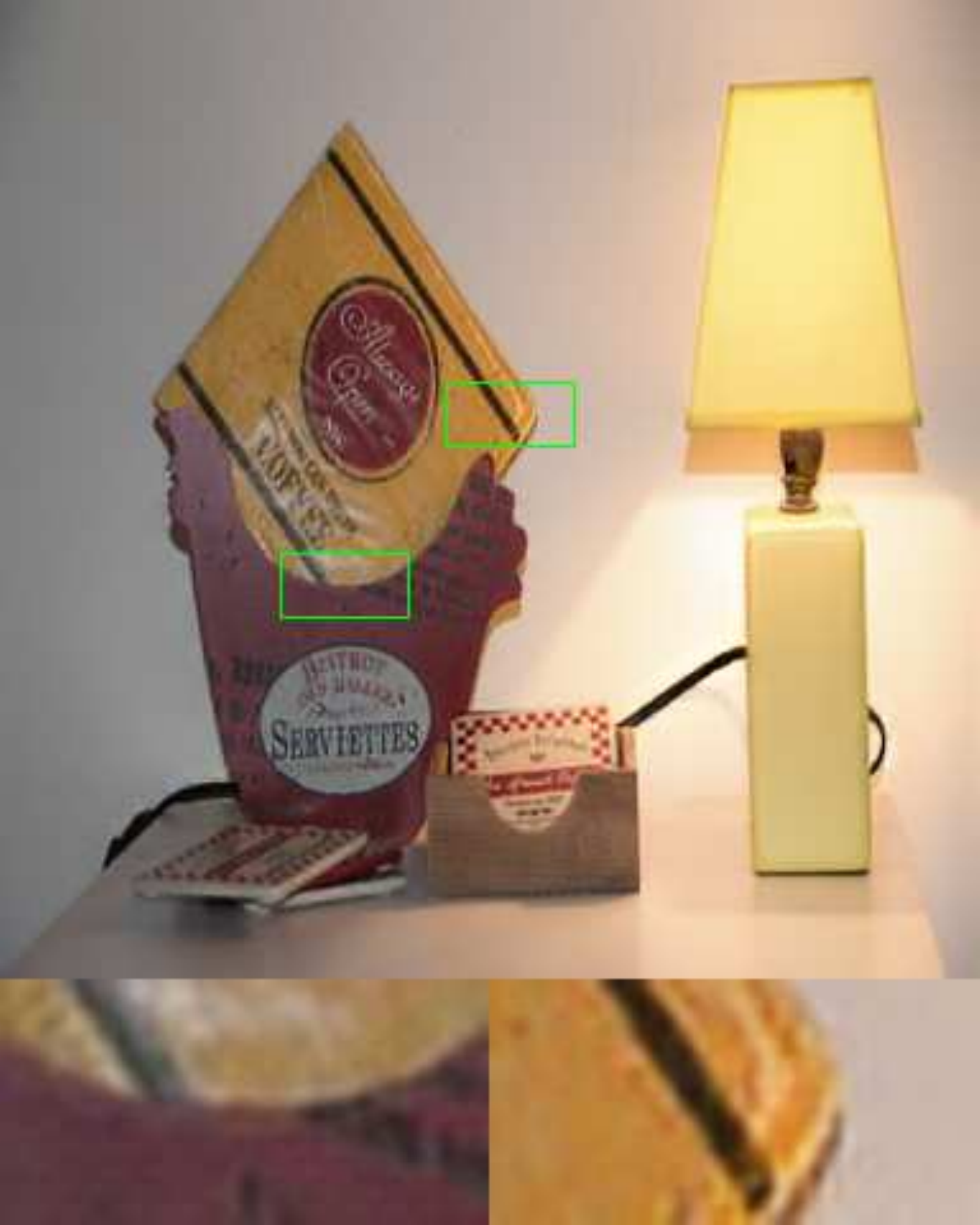}&\hspace{-5mm}
\\
\hspace{-4mm}
\includegraphics[width = 0.125\linewidth]{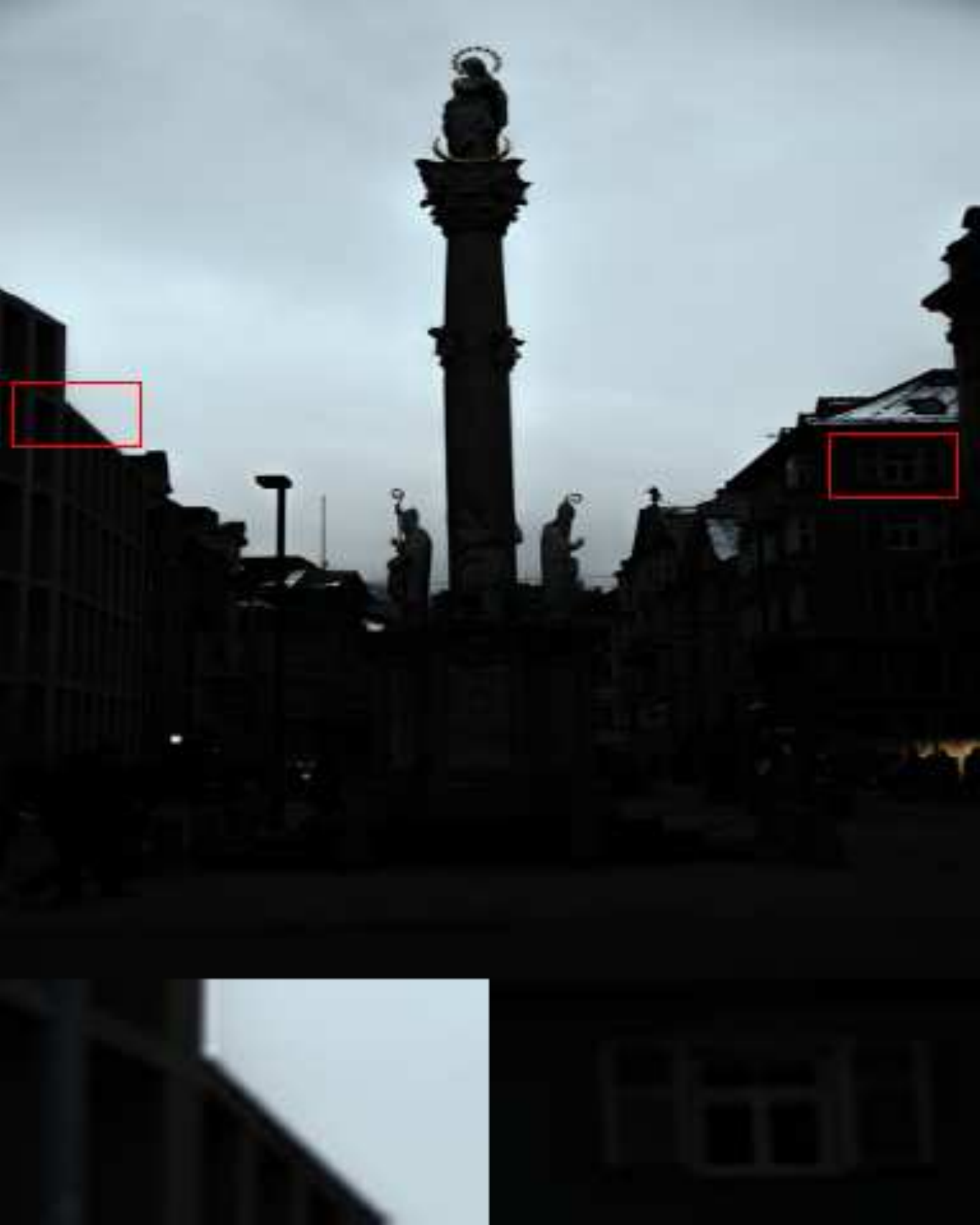} &\hspace{-5mm}
\includegraphics[width = 0.125\linewidth]{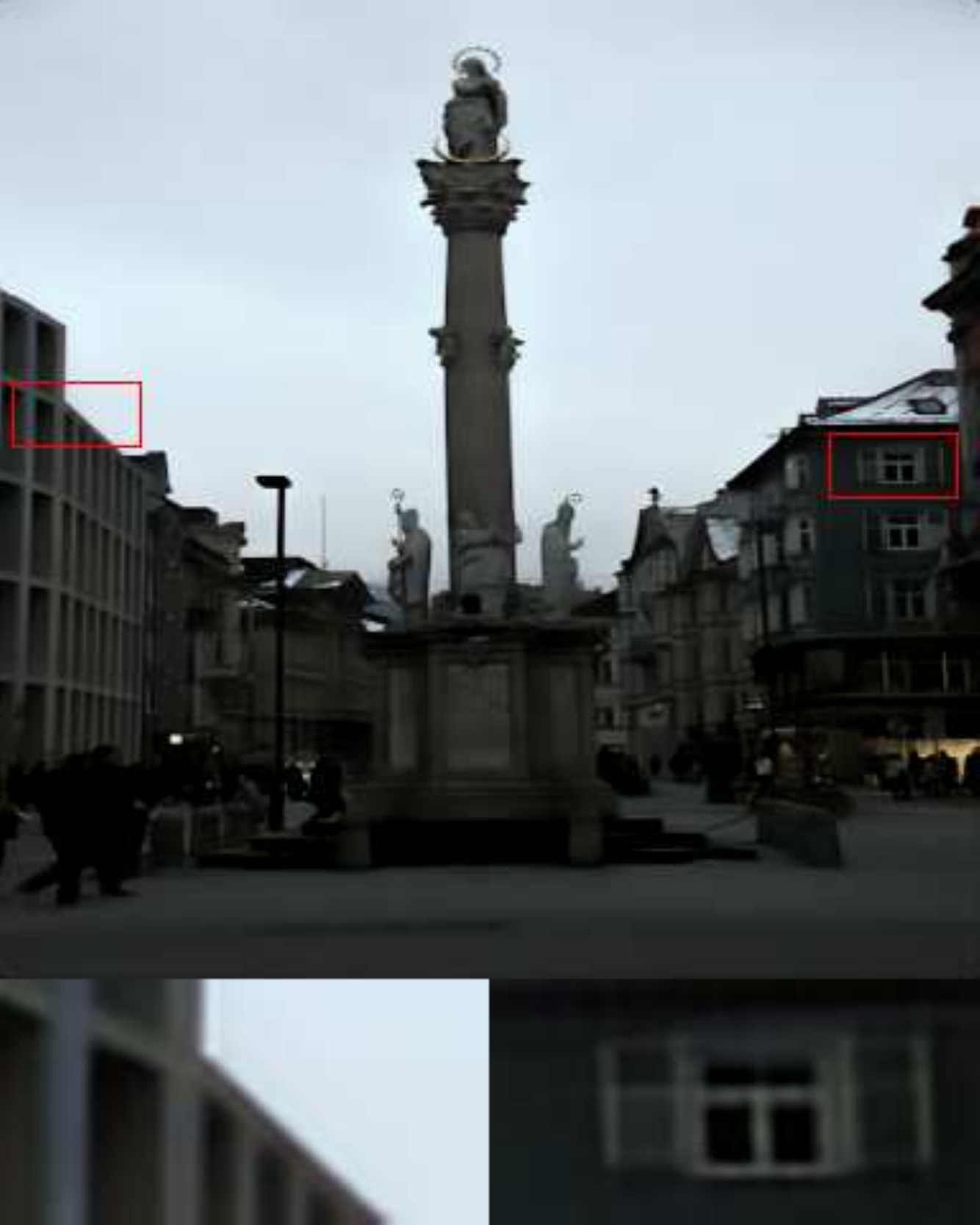}&\hspace{-5mm}
\includegraphics[width = 0.125\linewidth]{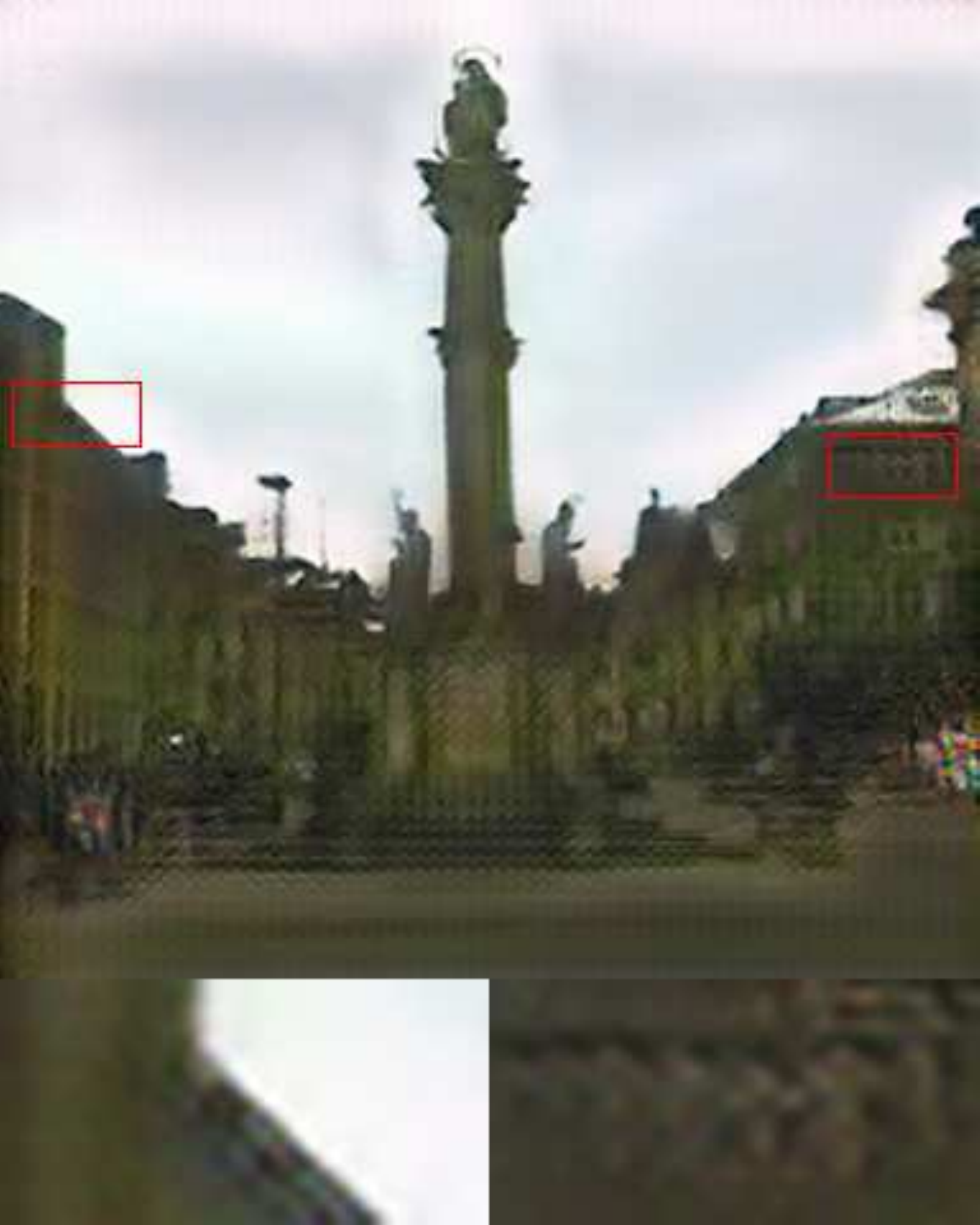}&\hspace{-5mm}
\includegraphics[width = 0.125\linewidth]{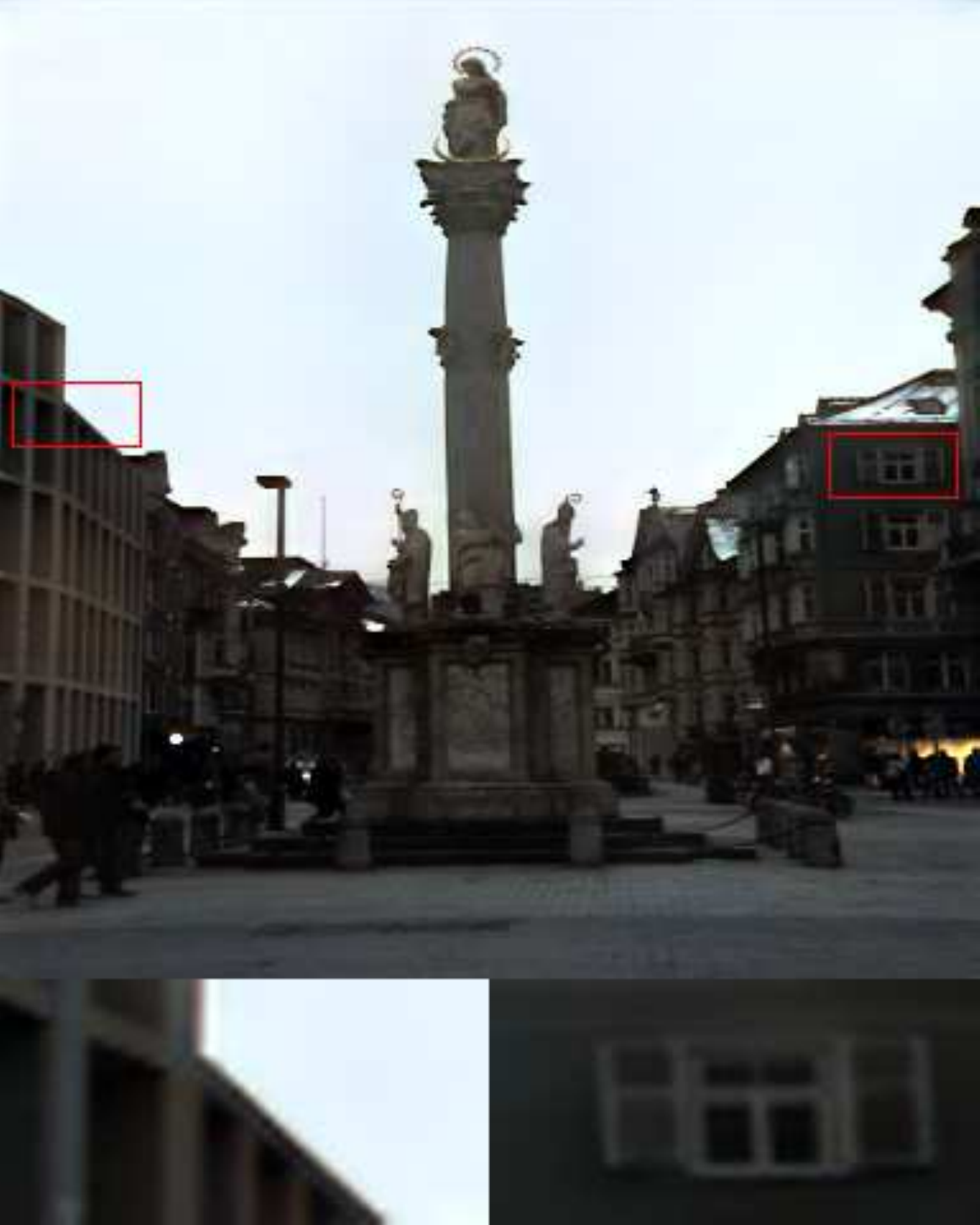}&\hspace{-5mm}
\includegraphics[width = 0.125\linewidth]{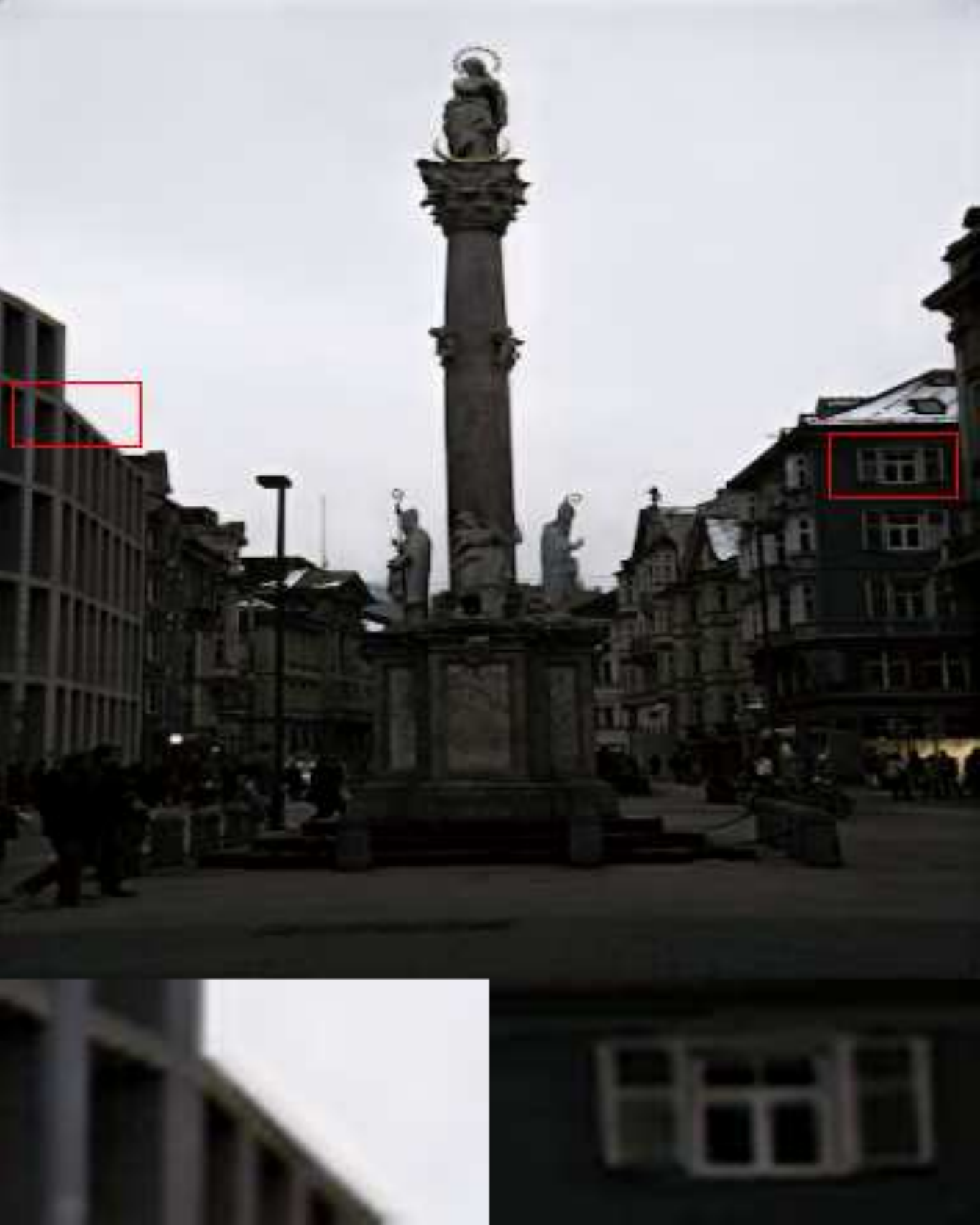}&\hspace{-5mm}
\includegraphics[width = 0.125\linewidth]{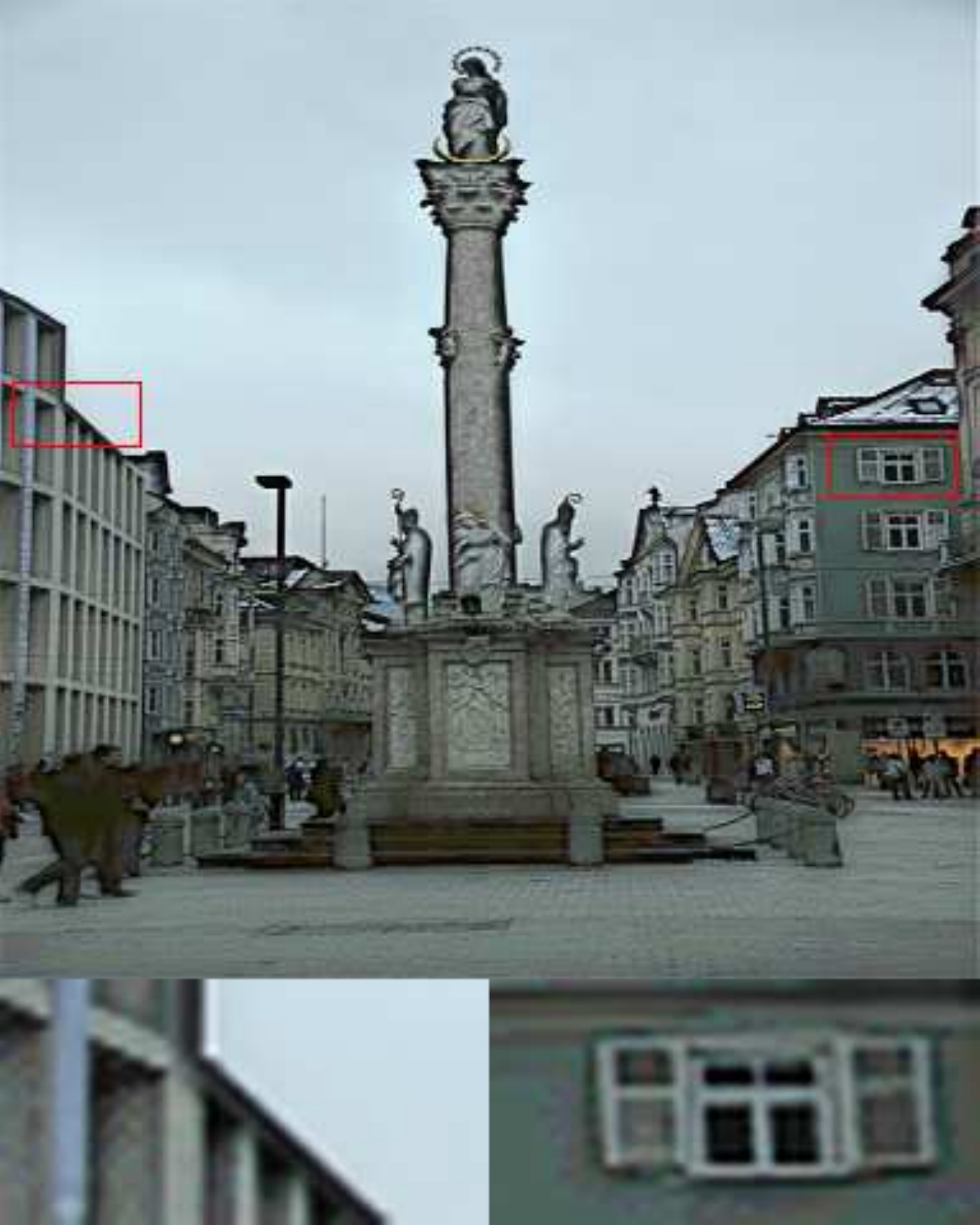}&\hspace{-5mm}
\includegraphics[width = 0.125\linewidth]{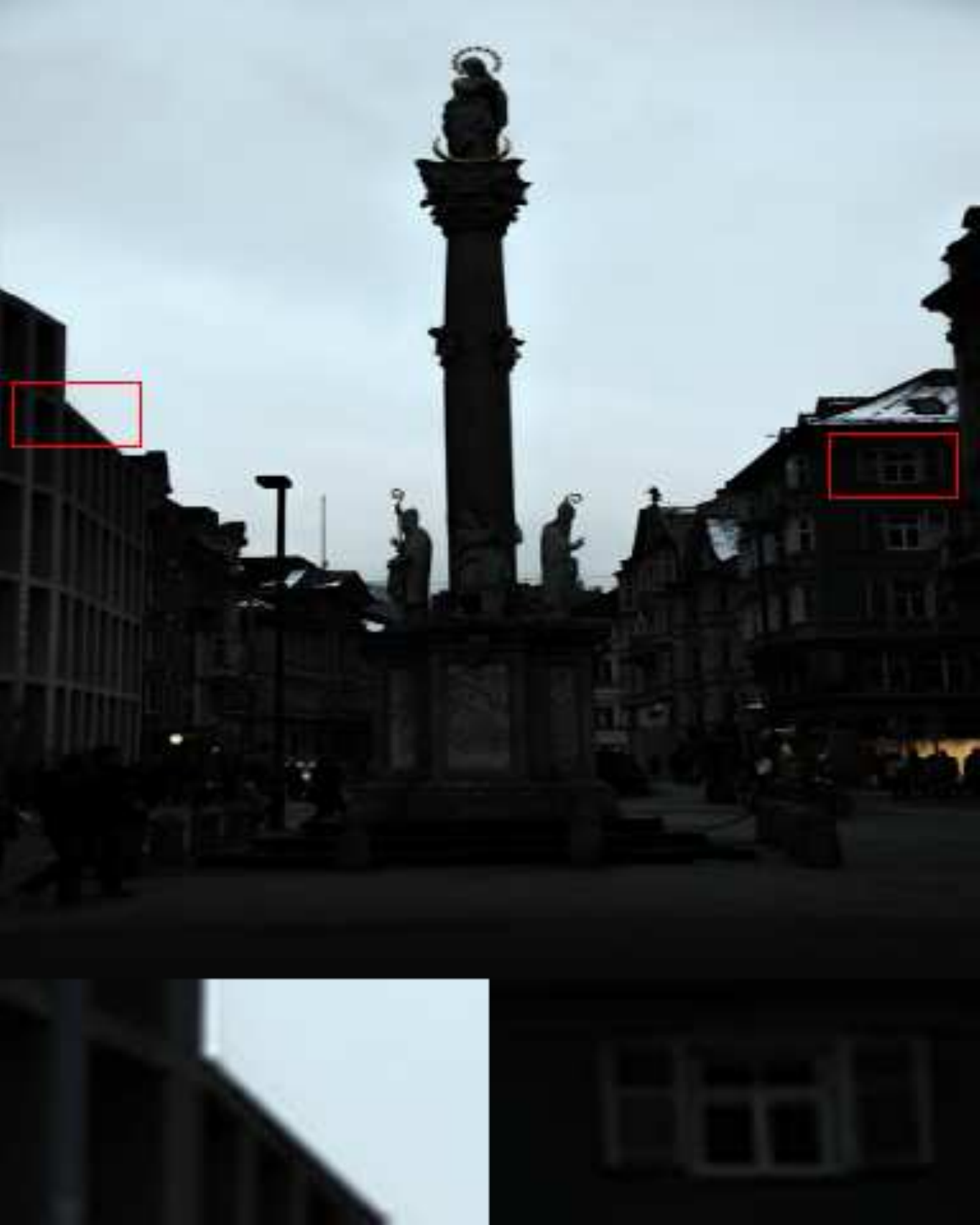}&\hspace{-5mm}
\includegraphics[width = 0.125\linewidth]{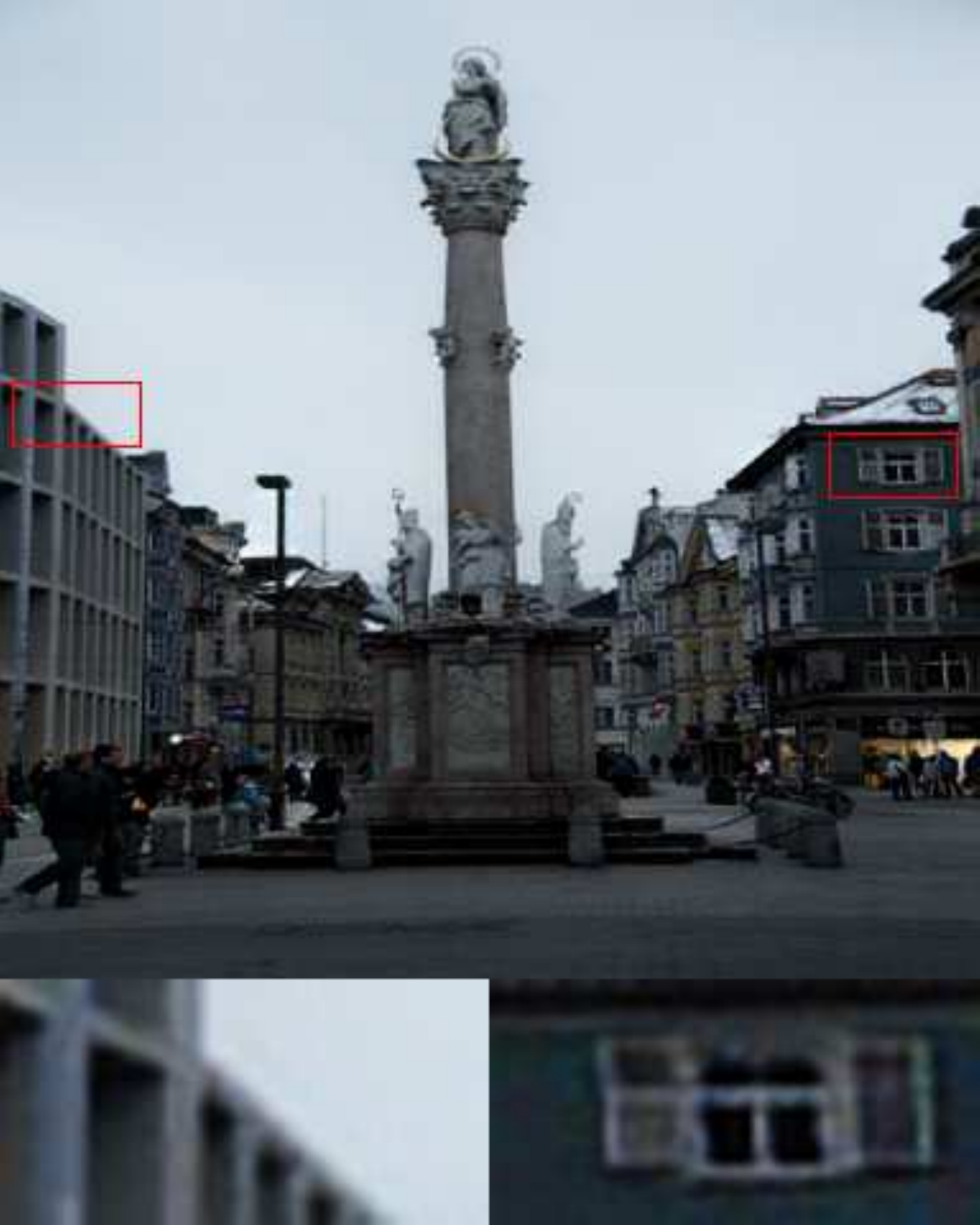}&\hspace{-5mm}
\\
Input &\hspace{-5mm}    MBLLEN  &\hspace{-5mm}  CycleGAN &\hspace{-5mm} EnlightenGAN &\hspace{-5mm} GLAD &\hspace{-5mm} RetinexNet &\hspace{-5mm} SRIE&\hspace{-5mm} UMLE
\\
\end{tabular}
\end{center}
\caption{ Visual comparison with state-of-the-art low-light image enhancement methods. Boxes indicate the obvious differences. It can be seen  that our results contain little noise and obtain the most natural color of the image.}
\label{fig:all}
\end{figure*}

\subsection{Generator}
Because  shared structure is adopted, the generator use the same encoder as discriminator. For the decoder, we also adopt the CPAM to extract features. After that, the model can obtain a nice  image by up-sampling.

\subsection{Loss function}
The  following five losses are adopted to train the UMLE model:
\subsubsection{Adversarial loss} We adopt the LSGAN \cite{lsgan} loss as the adversarial loss. This loss  matches the distribution of translated images and target images.

\begin{equation}
\begin{aligned}
\mathcal{L}_{adv}&=\mathbb{E}_{x \sim \mathcal{X}_{N}}\left[\log D_{N}(x)\right] \\
&+\mathbb{E}_{x \sim \mathcal{X}_{L}}\left[\log \left(1-D_{N}(G_{L\rightarrow N}(x)\right)]\right..
\end{aligned}
\end{equation}

\subsubsection{Cycle loss} This loss is adopted  to  make the image as similar as possible to the original image after two transformations.

\begin{equation}
\begin{aligned}
\mathcal{L}_{cyc} &=\mathbb{E}_{x \sim \mathcal{X}_{L}}\left[\|x-G_{N\rightarrow L}(G_{L\rightarrow N}(x))\|_{1}\right].
\end{aligned}
\end{equation}

\subsubsection{Color loss} Color loss forces the enhanced images to have similar color distributions as the target high-quality pictures.
\begin{equation}
\mathcal{L}_{\text {color }}=\mathbb{E}_{x \sim \mathcal{X}_{L}} [\log (D_{N}\left(G_{L\rightarrow N}(x)\right))] .
\end{equation}
\subsubsection{Preserving Loss} The function of preserving loss is to limit the vgg characteristic distance between the input low light level and its enhanced normal light output.
\begin{equation}
\mathcal{L}_{pre}=\frac{1}{W_{i} H_{i}} \sum_{x=1}^{W_{i}} \sum_{y=1}^{H_{i}}\left(\phi_{i}\left(x\right)-\phi_{i}\left(G_{L\rightarrow N}(x))\right)\right)^{2},
\end{equation}
where $\phi_{i}$ denotes the $i$-th feature map extracted from a pre-trained VGG-16 model; and  $H_{i}$ and $W_{i}$ are the height and width of the feature maps, respectively.

\subsubsection{Reconstruction loss} When the real sample of the source domain is provided as the input of the source domain generator, the transformed image need to be as similar to the real sample.

\begin{equation}
\mathcal{L}_{idt}=\mathbb{E}_{x \sim \mathcal{X}_{N}}\left[\left|x-G_{L\rightarrow N}(x)\right|_{1}\right] .
\end{equation}
The whole loss function of our model is as follows:
\begin{equation}
\mathcal{L}_{all}=\omega_1\mathcal{L}_{adv}+\omega_2\mathcal{L}_{cyc}+\omega_3\mathcal{L}_{color}+\omega_4\mathcal{L}_{pre}+\omega_5\mathcal{L}_{idt}.
\end{equation}
\par In the experiments, we empirically set $\omega_1$=1,  $\omega_2$=100, $\omega_3$=0.005, $\omega_4$=0.005, $\omega_1$=10.

\section{Experiments}
In this section, we first compare the performance of our method with state-of-the-art image enhancement methods to demonstrate the validity of our proposed method, which consists of experiments (A)$-$(F). In addition, we demonstrate the utility of our method by applying our method to SLAM relocation and drivable area detection, which are part of the experiments including (G), (H).
\subsection{Dataset}
In order to evaluate our method comprehensively,  we train it on various scenes including indoor and outdoor scenes. Since our method does not need paired dataset to train, we randomly select images of different scenes and different illuminations from existing datasets which released in \cite{62} , \cite{dataset2} and \cite{31}. The dataset includes 923 low light images and 750 normal light images.
\par For the relocalization test, we used the dataset published by ETH \cite{eth}.
\subsection{Implementation details}
We use ReLU \cite{relu} as the activation function. The generator and discriminator are trained by  AdamGC \cite{AdamGC} optimizer, and we  use a weight decay at the rate of 0.0001. For the normalization layer, the generator uses Adalin \cite{49} which combines the characteristics of adain \cite{54} and LN \cite{53} which can solve channel unrelated problems.  Due to memory constraints, we set the batch size  to 1 for our  experiments.  The training process for our model is carried out on an Nvidia P40. Because our model adopts the method of structure sharing, the training time of our model is greatly reduced. Our models are trained using only 50K iterations.

%\begin{table*}[t]
%\caption{Quantitative evaluation of low-light image enhancement algorithms.}
%\begin{center}
%\label{table:1}
%\begin{tabular}{|c||c||c||c||c||c||c||c|}
%\hline
%Model & Retinex-Net& MBLLEN& GLADNet& CycleGAN& EnlightenGAN&SRIE& UMLE  \\[3pt]
%\hline
%NIQE($\downarrow$)& 7.539 & 4.481 & {\color{red}\textbf{3.254}}& 4.260& \textbf{3.572}& 4.659& {\color{blue}\textbf{3.485}}\\[3pt]
%ENTROPY($\uparrow$) & 6.923 & 7.050 &  \textbf{7.340} & 7.092&  {\color{blue}\textbf{7.350}}& 6.874&  {\color{red}\textbf{7.506}}\\[3pt]
%\hline
%%\multicolumn{5}{p{230pt}}{Bold blue represents the best result, and bold black represents the second result. Larger Entropy and Vollath means higher definition, and smaller PCA-based and DRL indicates less image noise.}
%
%\end{tabular}
%
%\end{center}
%\end{table*}

\begin{table}[h]{\centering}
\caption{Quantitative evaluation of low-light image enhancement algorithms.}
\label{table:1}
\begin{center}
\begin{tabular}{|c||c||c|}
\hline
Model  & NIQE($\downarrow$) & ENTROPY($\uparrow$)\\[3pt]
\hline
Retinex-Net & 7.539 & 6.923 \\[3pt]
\hline
MBLLEN & 4.481 & 7.050  \\[3pt]
\hline
GLADNet & {\color{red}\textbf{3.254}} &  \textbf{7.340}\\[3pt]
\hline
CycleGAN & 4.260 & 7.092 \\[3pt]
\hline
EnlightenGAN & \textbf{3.572}& {\color{blue}\textbf{7.350}}   \\[3pt]
\hline
SRIE &  4.659 & 6.874 \\[3pt]
\hline
UMLE & {\color{blue}\textbf{3.485}} & {\color{red}\textbf{7.506}}\\[3pt]
\hline
\end{tabular}
\end{center}
\end{table}

\begin{figure}[t]
\centering
\includegraphics[width=0.48\textwidth]{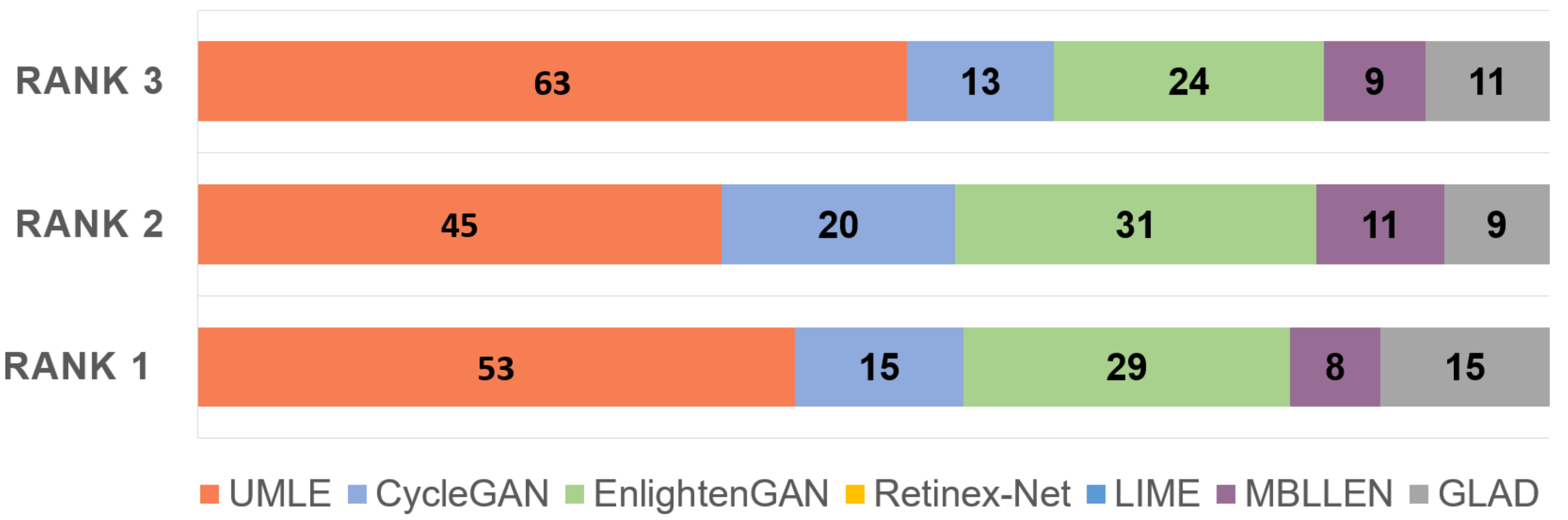} % Reduce the figure size so that it is slightly narrower than the column.
\caption{User study score on the test dataset, the number on the picture means the distribution of the images that the user thinks is the best for each rank.}
\label{fig:rank}
\end{figure}

\subsection{Qualitative Comparison}
We conduct extensive qualitative evaluations on typical low-light images with several state-of-art methods. The results are illustrated in Fig.~\ref{fig:all}. with MBLLEN \cite{MBLLEN}, Cyclegan \cite{46}, EnlightenGAN \cite{31}, GLAD \cite{GLAD}, Retinex-Net \cite{62}, SRIE \cite{7}.
In terms of color and brightness, Enlightengan \cite{31} is prone to color deviation and the result of SRIE \cite{7} is likely to be darker. For texture, LIME \cite{10} generates  more noise while enhancing. The results of Cyclegan \cite{46} have the problem of fuzzy details.  Retinex-Net \cite{62} produces a kind of effect similar to crayon drawing, which affects the quality of generation. In contrast, our method performs well in terms of texture and color.

\subsection{User study}
We adopt a human subjective evaluation \cite{31} which is a human subjective study to compare the performance of our method and other methods. This approach is similar to the double stimulus continuous scale. It evaluates the image quality from the following three indicators:
\par Rank1: How much visible noise the images contain;
\par Rank2: How much over or under exposure artifacts the images contain;
\par Rank3: Whether the images show nonrealistic color or texture distortions.
\par We randomly selected 120 volunteers to evaluate the results. For each indicator, the distributions of the best performing images in all methods are shown in Fig.~\ref{fig:rank}. The number of images that each method performs best in for each category is included in the figure.
\par By analyzing these results, it can be seen that our method performs best in the three indicators, and our result is natural in color with almost no visible noise.

\begin{table}[t]
\caption{The number of parameters for generator, training iterations and testing FPS of each model}
\label{size}
\begin{tabular}{|c||c||c||c|}
\hline
\diagbox{Model}{Module} &  Parameters  &  Training iterations & FPS \\
\hline
EnlightenGAN & 22.30M & 327k & 52\\[3pt]
\hline
CycleGAN &{\textbf{11.37M}} & 327k & 45\\[3pt]
\hline
ULEN(Our) & 16.19M & {\textbf{50k}} & {\textbf{65}}\\[3pt]
\hline
\end{tabular}
\end{table}

\begin{figure}[t]
\begin{center}
\begin{tabular}{ccccccc}
\hspace{-3mm}
\includegraphics[width = 0.16\linewidth]{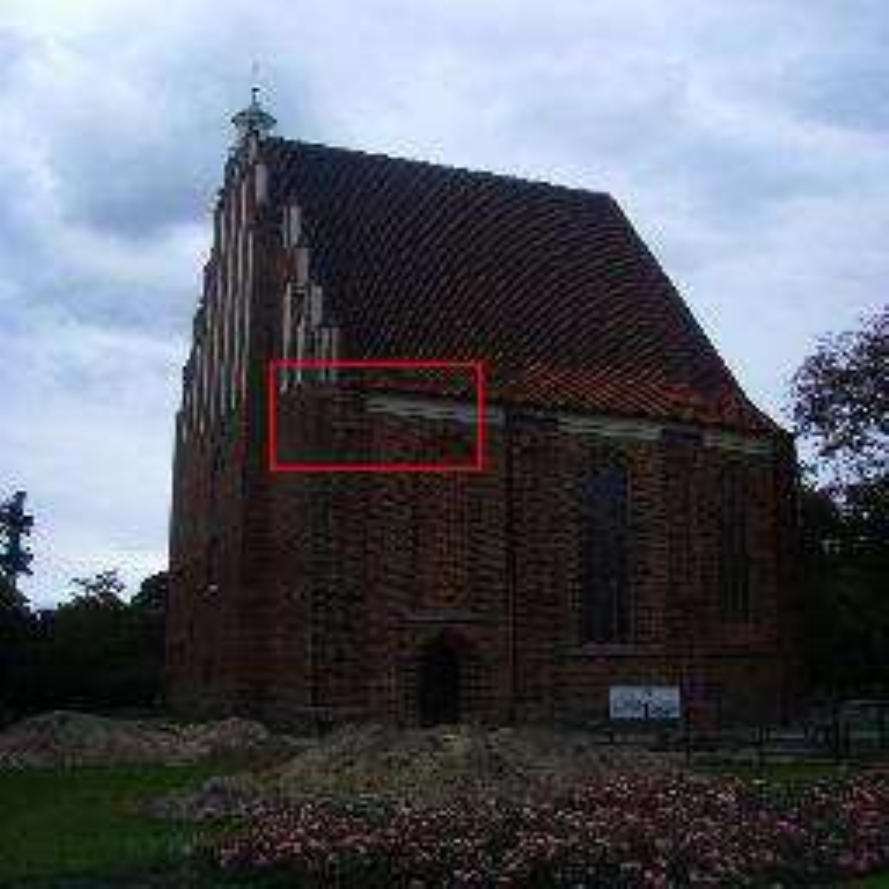} &\hspace{-5mm}
\includegraphics[width = 0.16\linewidth]{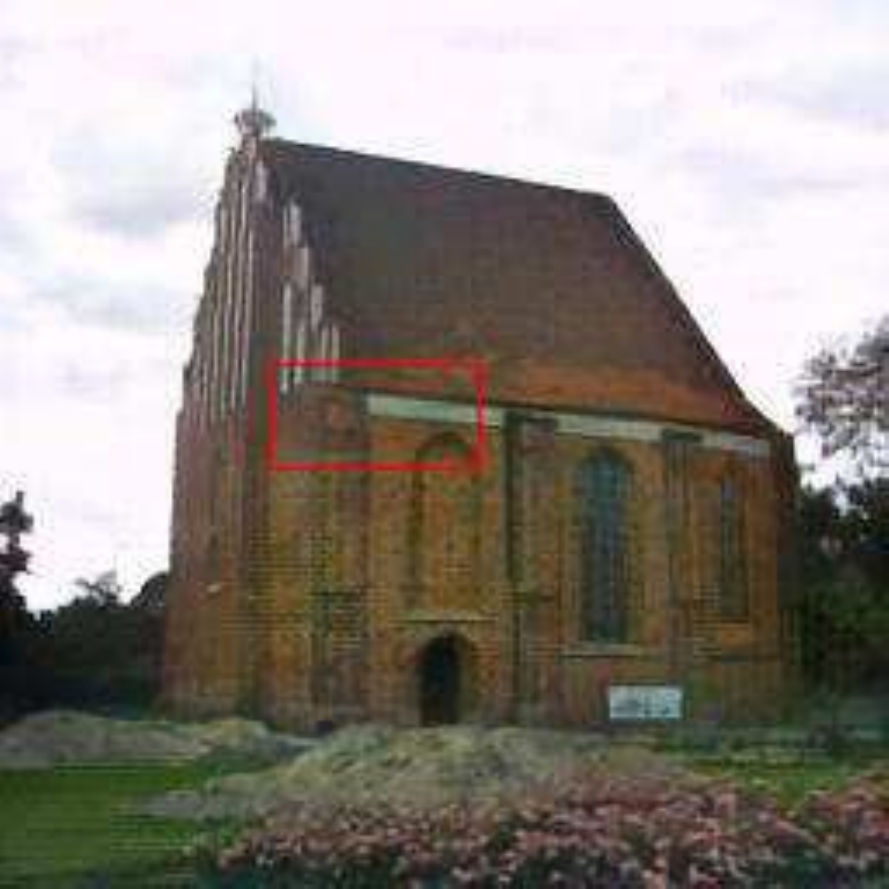}&\hspace{-5mm}
\includegraphics[width = 0.16\linewidth]{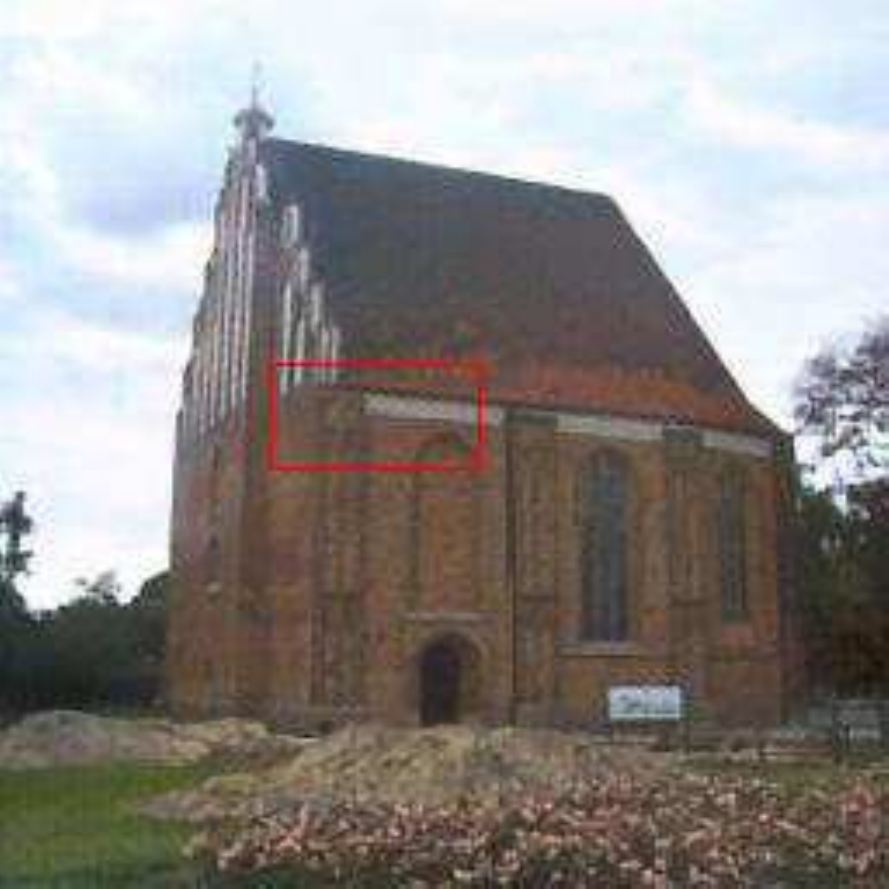}&\hspace{-5mm}
\includegraphics[width = 0.16\linewidth]{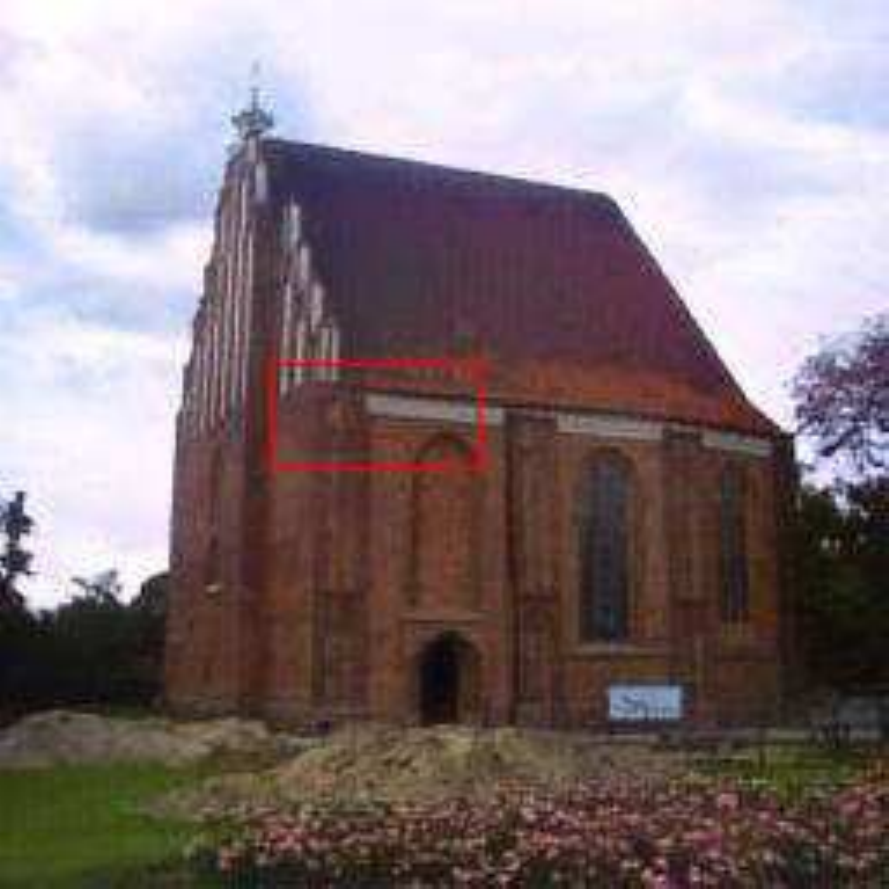}&\hspace{-5mm}
\includegraphics[width = 0.16\linewidth]{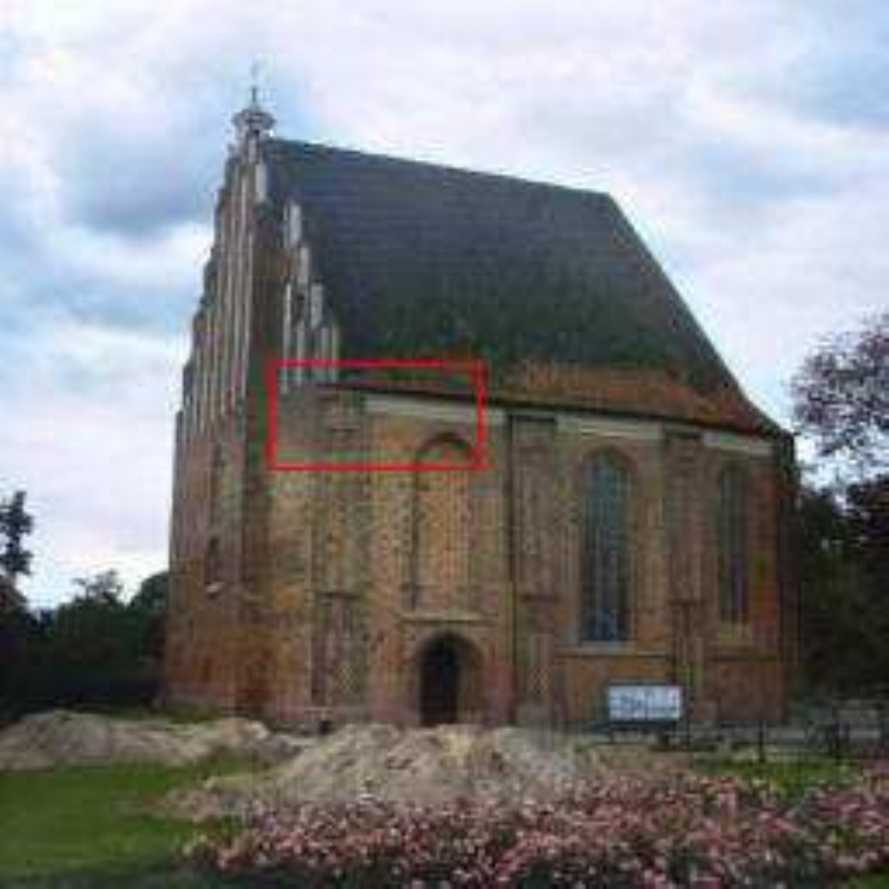}&\hspace{-5mm}
\includegraphics[width = 0.16\linewidth]{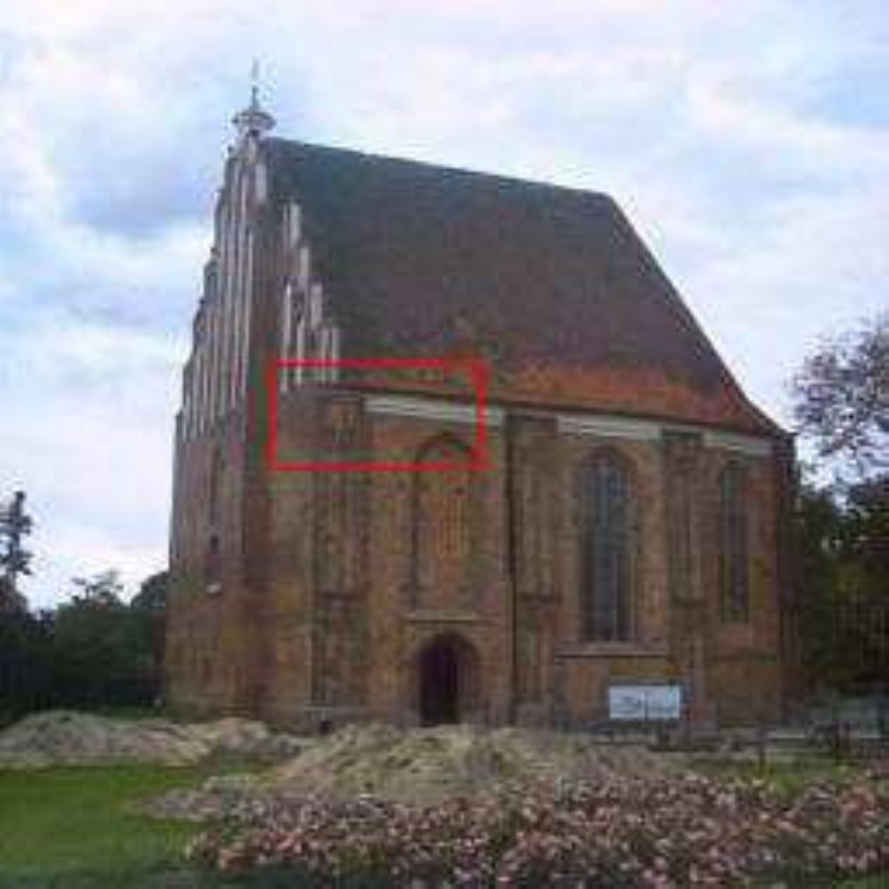}&\hspace{-5mm}
\\
\hspace{-3mm}
\includegraphics[width = 0.16\linewidth]{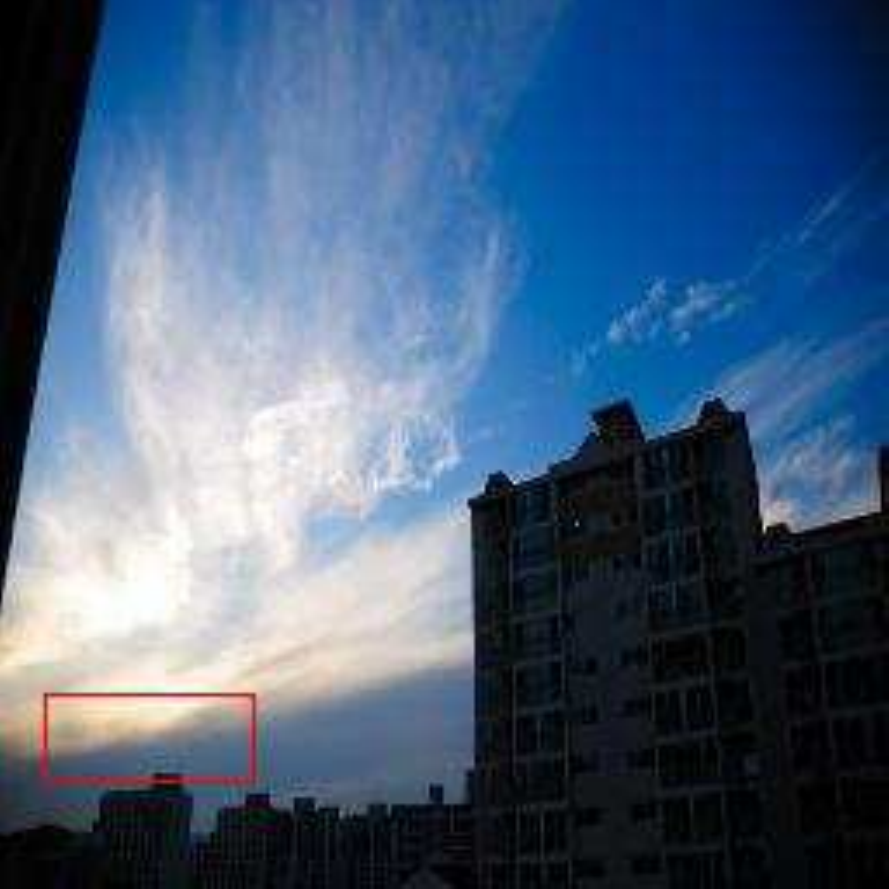} &\hspace{-5mm}
\includegraphics[width = 0.16\linewidth]{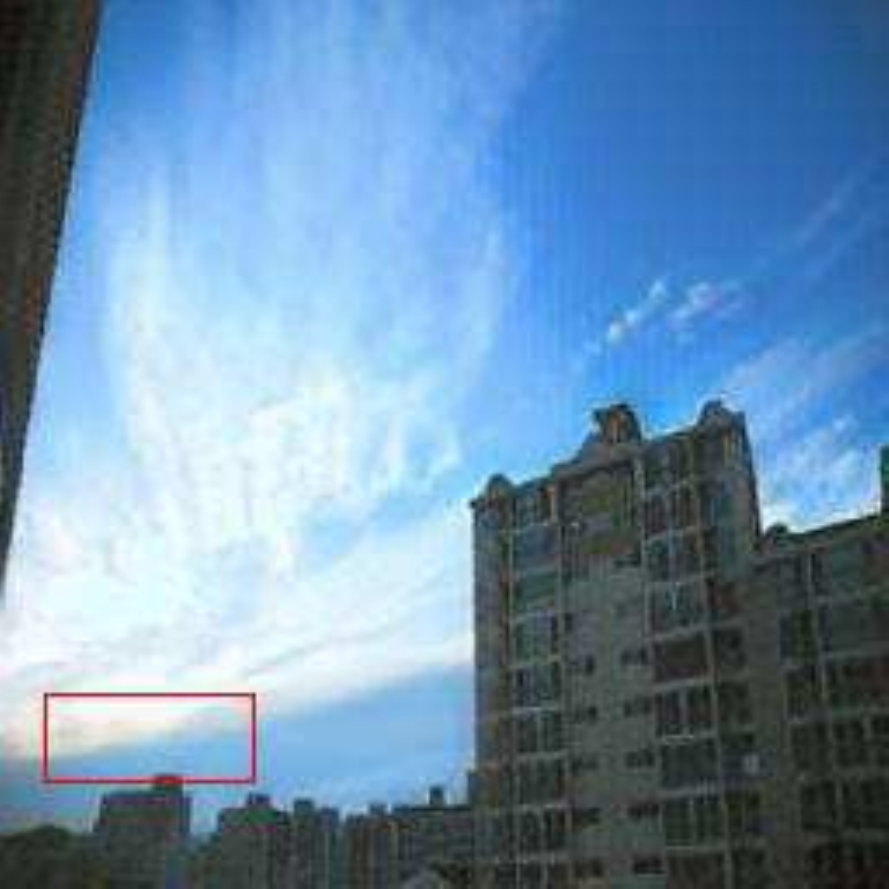}&\hspace{-5mm}
\includegraphics[width = 0.16\linewidth]{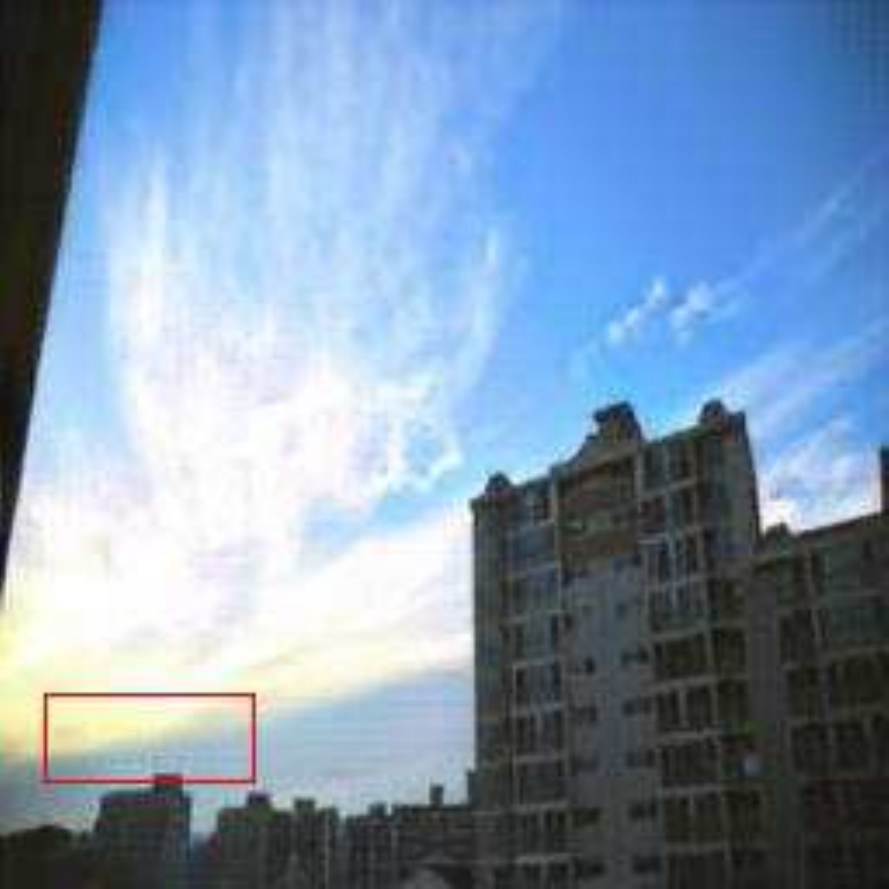}&\hspace{-5mm}
\includegraphics[width = 0.16\linewidth]{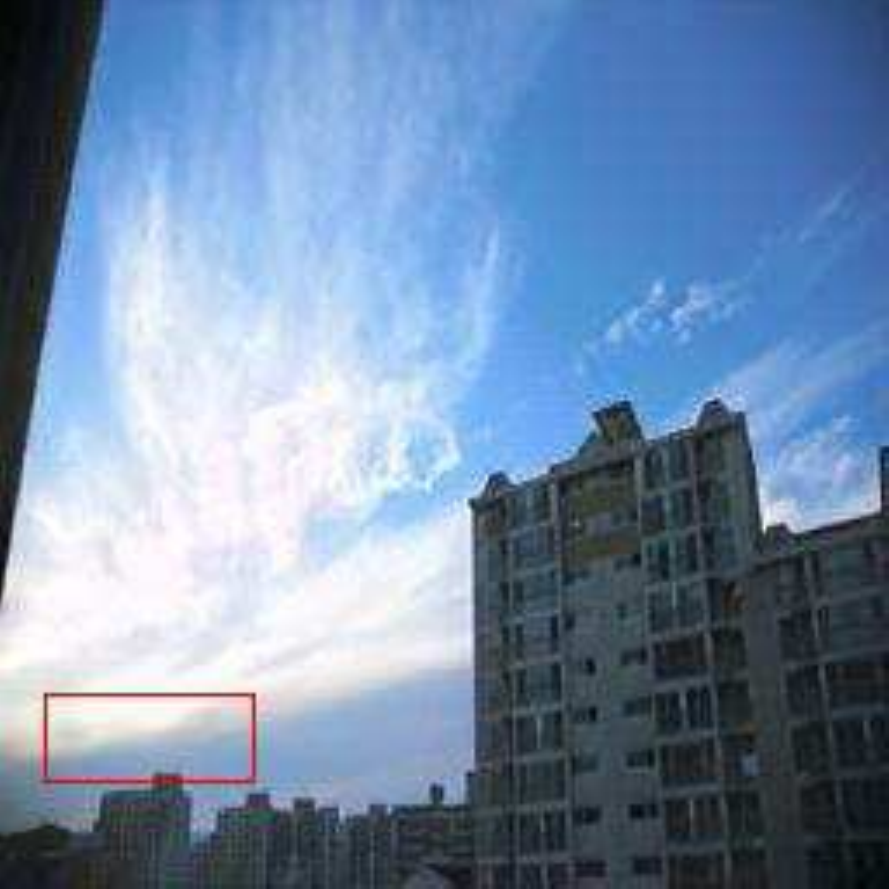}&\hspace{-5mm}
\includegraphics[width = 0.16\linewidth]{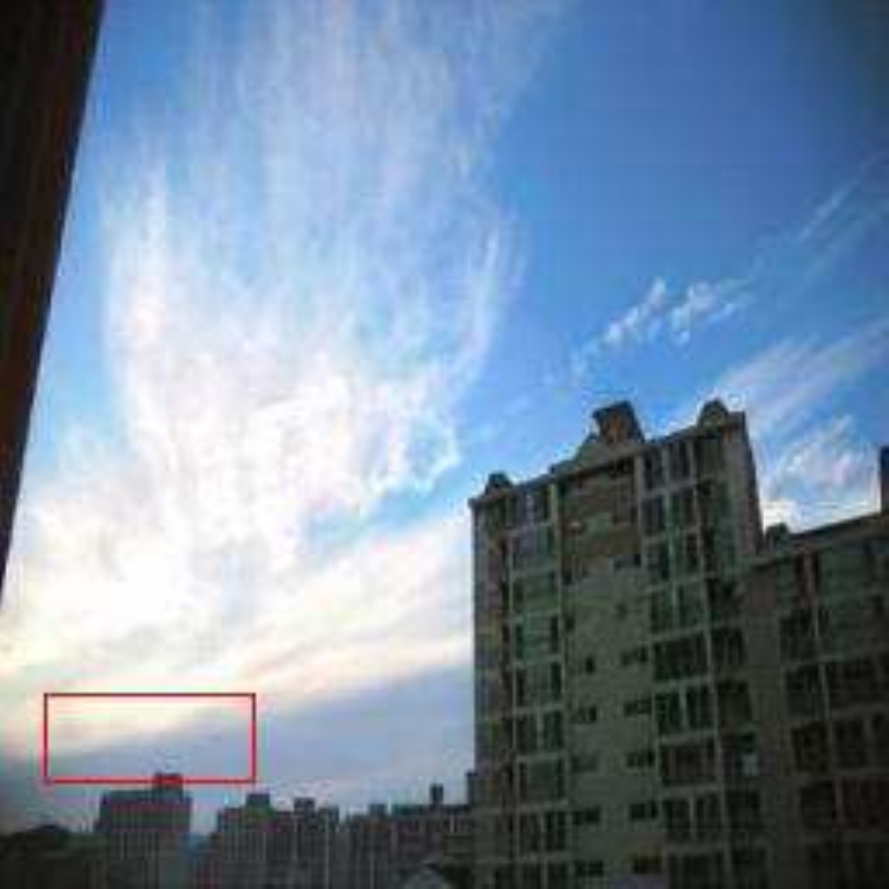}&\hspace{-5mm}
\includegraphics[width = 0.16\linewidth]{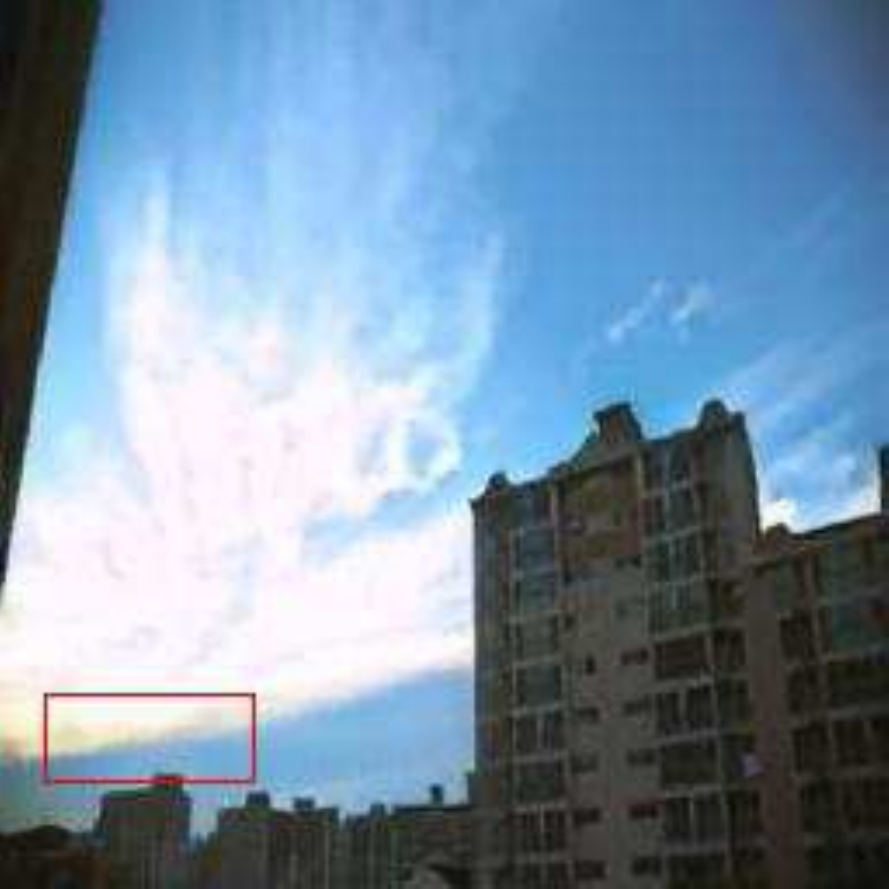}&\hspace{-5mm}
\\
Input &\hspace{-5mm} (a)  &\hspace{-5mm}  (b) &\hspace{-5mm} (c) &\hspace{-5mm} (d) &\hspace{-5mm} (e) &\hspace{-5mm}
\\
\end{tabular}
\end{center}
\caption{ Visual comparison with ablation results, boxes indicate the obvious differences. (a) is the result without color discriminator, (b) is the result without texture discriminator, (c) is the result without multi-scale discriminator, (d) is the result without CPAF module, (e) is the result of UMLE.}
\label{fig:duibi}
\end{figure}
\begin{table}[h]{\centering}
\caption{The result of ablation study.}
\label{ablation}
\begin{center}
\begin{tabular}{|c||c||c|}
\hline
Condition  & NIQE($\downarrow$) & ENTROPY($\uparrow$)\\[3pt]
\hline
with $D_C$, w/o $D_{T}$, w/o $D_{M}$ & 5.416 & 7.083 \\[3pt]
\hline
with $D_T$, w/o $D_{C}$, w/o $D_{M}$ & 5.656 & 6.910  \\[3pt]
\hline
with $D_M$, w/o $D_{T}$, w/o $D_{C}$ & 4.840 &  6.743\\[3pt]
\hline
with $D_{T}$, with $D_{M}$, w/o $D_C$ & 4.589 & 7.132 \\[3pt]
\hline
with $D_{C}$, with $D_{M}$, w/o $D_T$ & 3.789 & 7.291   \\[3pt]
\hline
with $D_{T}$, with $D_{C}$, w/o $D_M$ & 3.929 & 6.992 \\[3pt]
\hline
with $D_M$, $D_{T}$, $D_{C}$, w/o $CPAF$ & 6.625 & 7.067 \\[3pt]
\hline
default configuration & 3.485 & 7.506   \\[3pt]
\hline
\end{tabular}
\end{center}
\end{table}

\subsection{No-reference evaluation}
We employ  naturalness image quality evaluator (NIQE) \cite{NIQE} and entropy \cite{entropy} to evaluate the perceptual quality of the state-of-art methods. The NIQE is based on the construction of a series of features used to measure image quality, and lower NIQE \cite{NIQE} values reflect higher quality. The entropy reflects the amount of information carried by the image. The greater the information entropy of the image means the better the quality.  Tab.~\ref{table:1} shows the calculated results of our NIQE \cite{NIQE} and entropy \cite{entropy} indices. The red result represents the best, blue represents the second best, and bold black represents the third best result. From the table we can see that the proposed UMLE shows well performance in terms of NIQE \cite{NIQE} and entropy \cite{entropy}, and it further demonstrates the advantages of our model in generating normal light images.
\par In addition, we compare current unsupervised models from the the number of parameters, model training perspective and runtime. Tab.~\ref{size} shows the number of training iterations, generator parameters, and frames per second (FPS) in testing for CylceGAN  \cite{46}, EnlightenGAN  \cite{31}, and our method UMLE. All parameters are calculated according to the original article, while we convert the epoch provided to iteration, and parameters means generator parameters, iterations indicates the training iterations and frames per second. Bold black represents the best results. We can  see that our method has great advantages in training time, the total number of  parameters and testing FPS.

\begin{figure}[t]
\centering
\includegraphics[width=1\columnwidth]{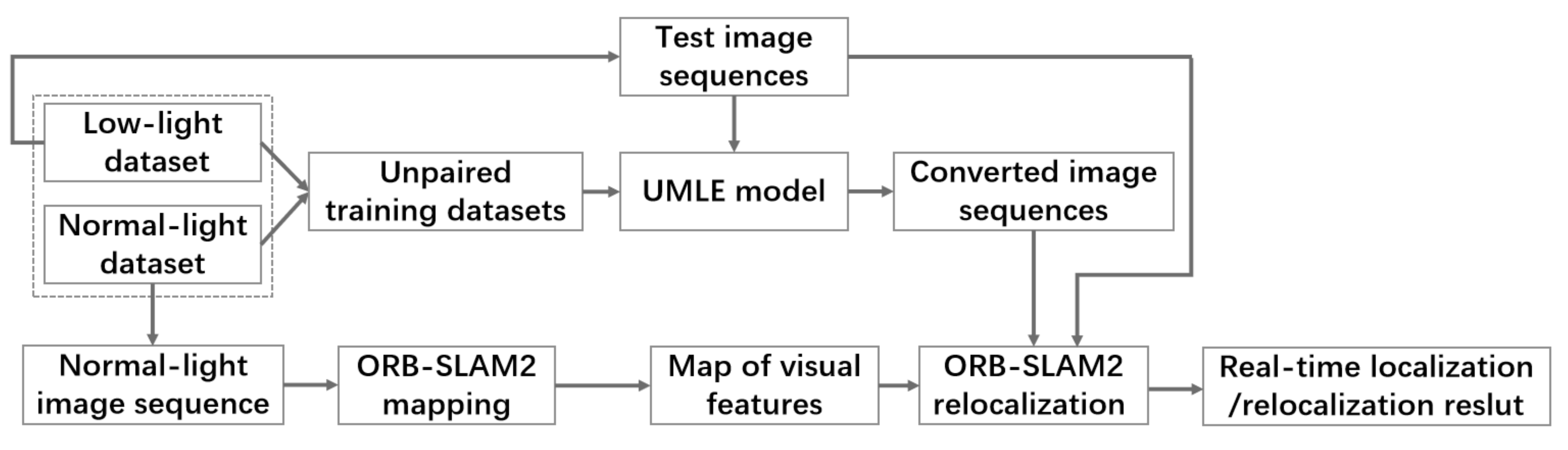} % Reduce the figure size so that it is slightly narrower than the column. Don't use precise values for figure width.This setup will avoid overfull boxes.
\caption{Flowchart of our experiment. We use the normal illumination image for mapping, while training the UMLE model using the unpaired flashlight illumination image and the normal illumination image. We then test the relocalization on the normal illumination build map using the enhanced image and the original flashlight illumination image, respectively.}
\label{liuchengtu}
\end{figure}

\subsection{Ablation study}
To verify  the effectiveness of each component proposed in Sec.3, we perform several ablation studies.
\par We tested the results separately in the absence of each discriminator. In Tab.~\ref{ablation} we calculate the NIQE \cite{NIQE} value in each case and verify experimentally the importance of each component. The results show that each module contributes substantially to improve the image quality.

\begin{figure}[t]
\centering
\includegraphics[width=1\columnwidth]{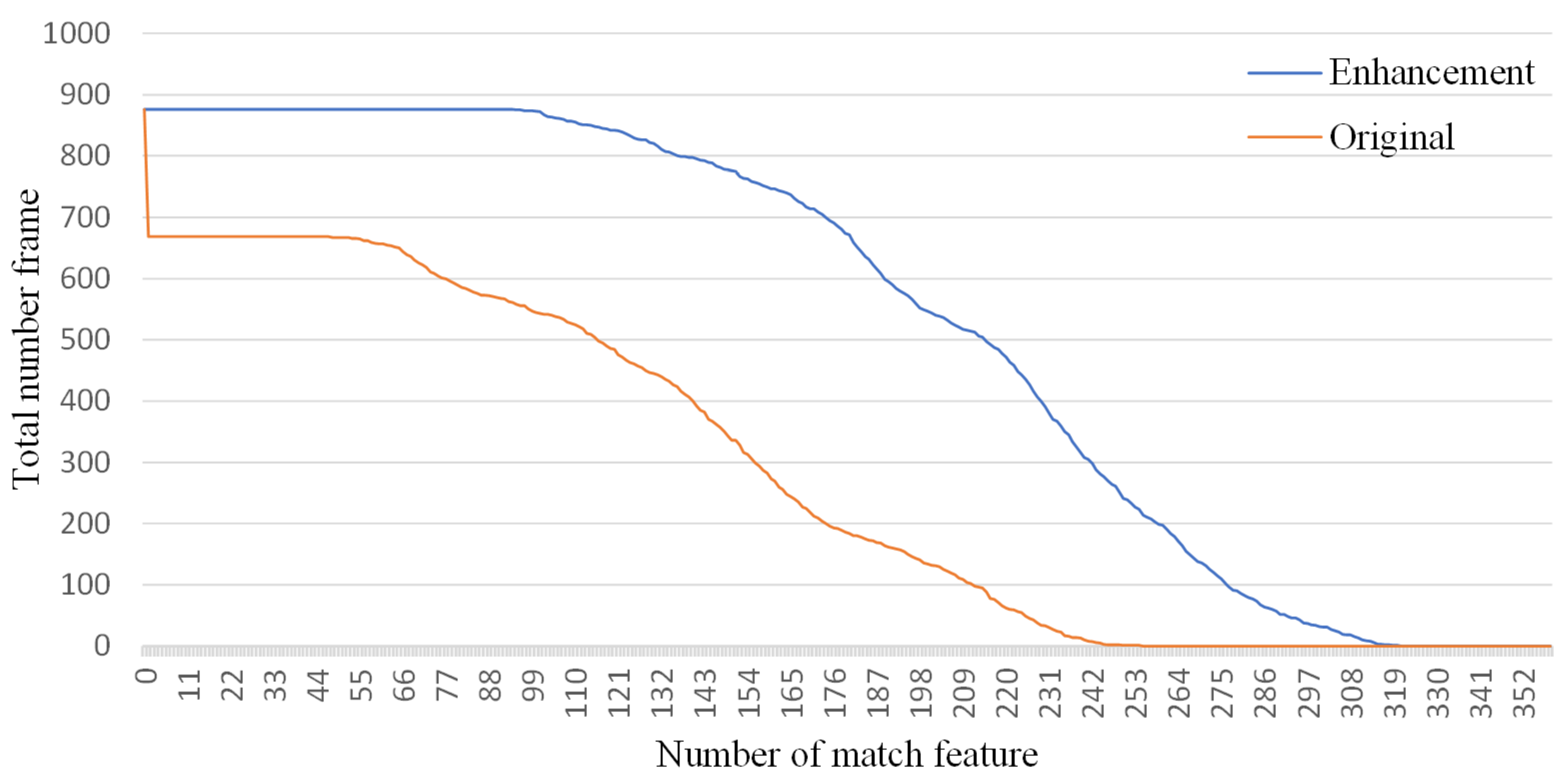} % Reduce the figure size so that it is slightly narrower than the column. Don't use precise values for figure width.This setup will avoid overfull boxes.
\caption{Comparison of the number of successfully matched feature points before and after image enhancement. The horizontal coordinate represents the number of successful matches, and the vertical coordinate represents the number of images with more features than the corresponding horizontal coordinate. From the image, we can see that the number of features that can be matched in the enhanced image is significantly increased.}
\label{test1}
\end{figure}

\begin{figure}[t]
\begin{center}
\begin{tabular}{ccccc}
\hspace{-3.5mm}
\includegraphics[width = 0.49\linewidth]{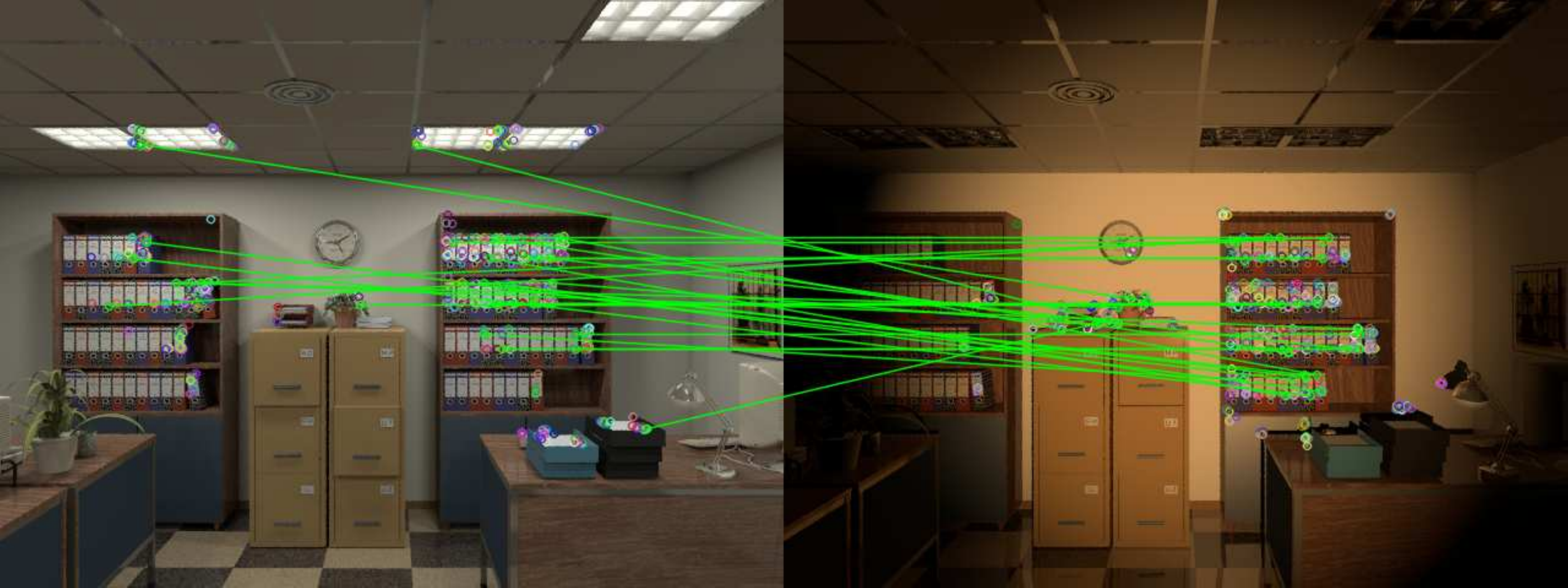} &\hspace{-5mm}
\includegraphics[width = 0.49\linewidth]{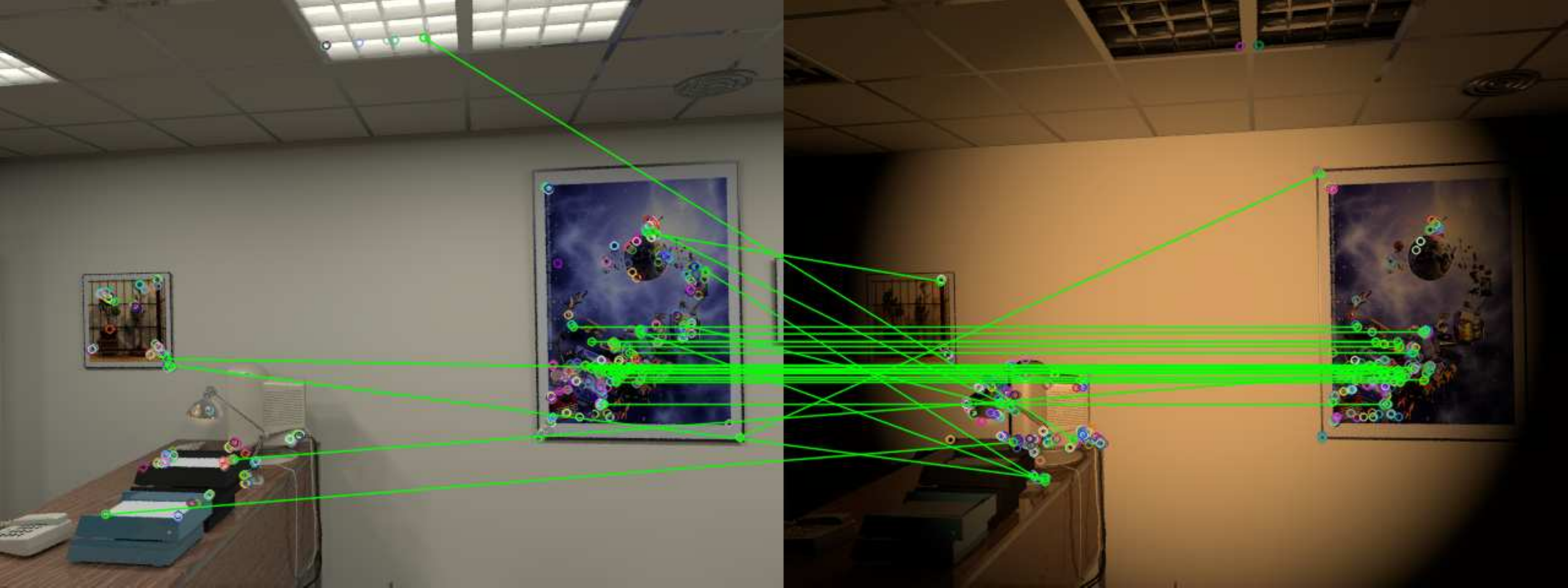} &\hspace{-5mm}
\\
\hspace{-3.5mm}
\includegraphics[width = 0.49\linewidth]{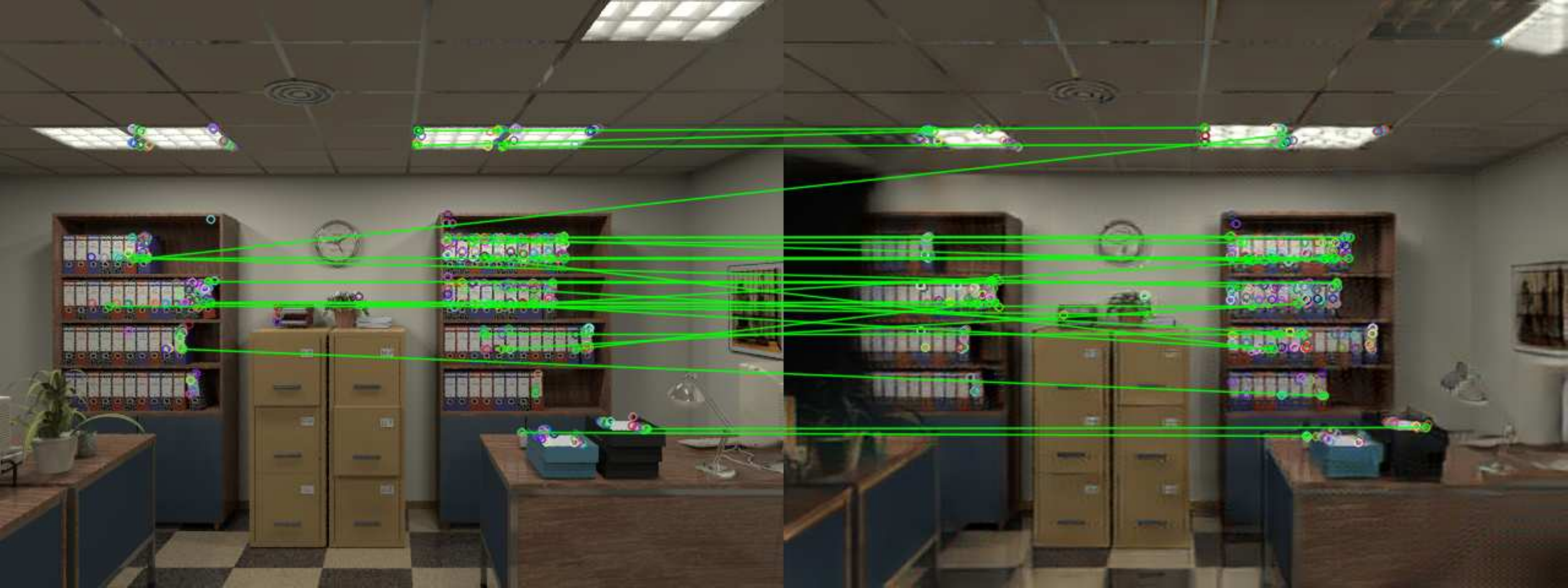} &\hspace{-5mm}
\includegraphics[width = 0.49\linewidth]{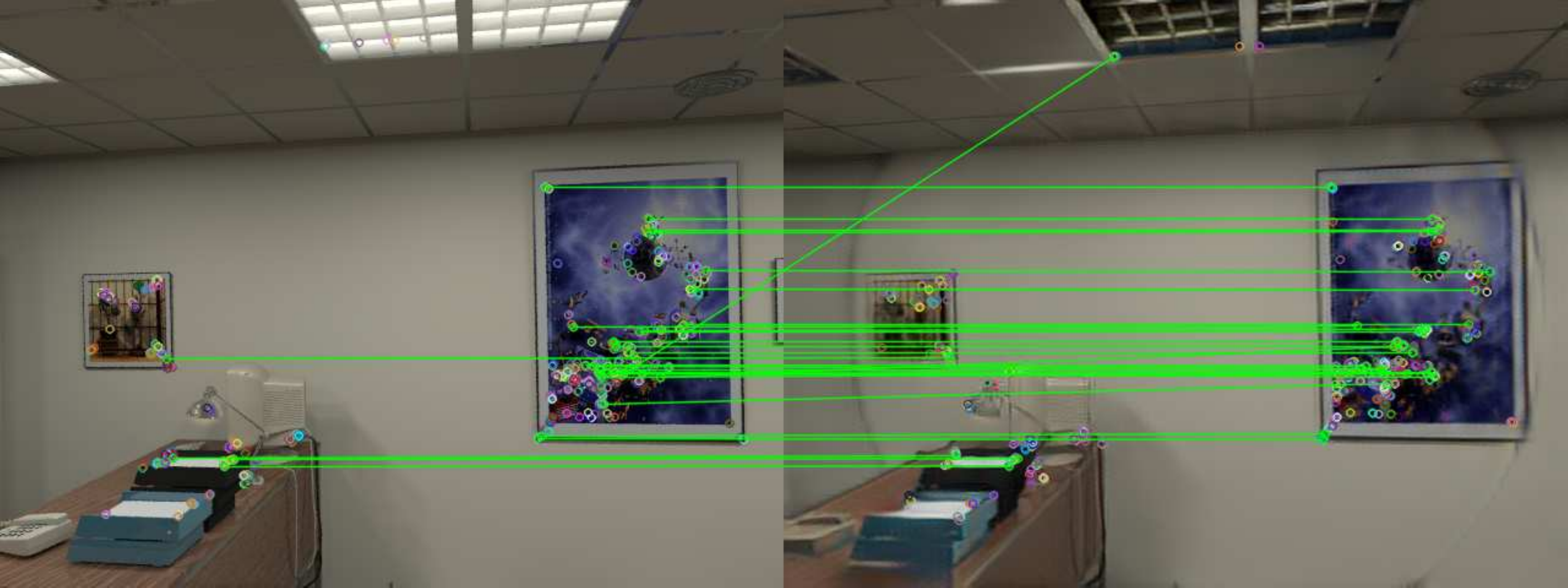} &\hspace{-5mm}
\\
(a) &\hspace{-4mm} (b)
\\
\end{tabular}
\end{center}
\caption{ ORB matching results before and after image enhancement. The first line is the result of matching the flashlight image with the normal light image, and the second line is the result of matching the enhanced image with the normal light image after the UMLE enhancement. We can find that the enhanced image matching results have significant advantages.}
\label{orbres}
\end{figure}

\begin{figure}[t]
\centering
\includegraphics[width=1\columnwidth]{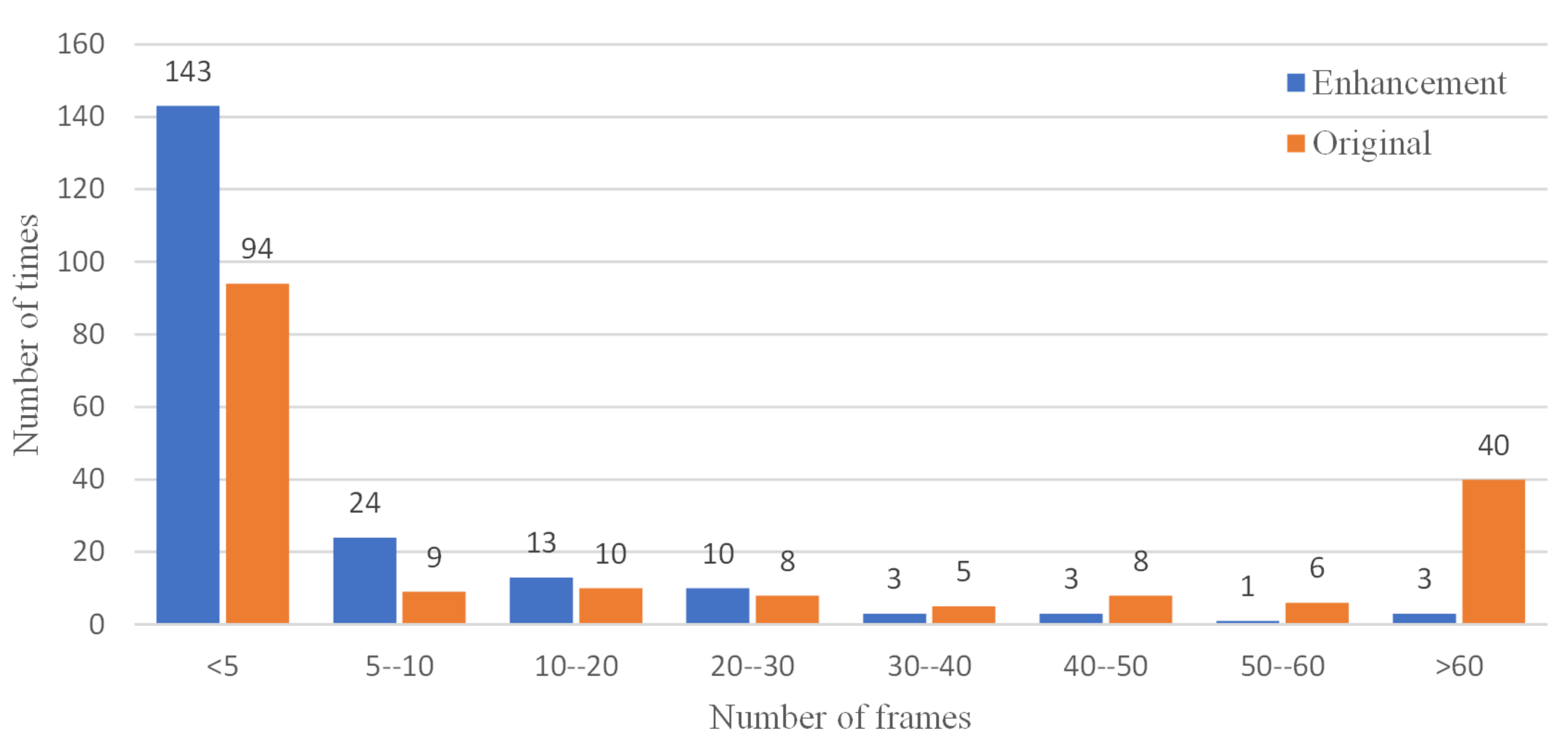} % Reduce the figure size so that it is slightly narrower than the column. Don't use precise values for figure width.This setup will avoid overfull boxes.
\caption{The number of images required for successful repositioning. The meaning of the horizontal coordinate in the image is the number of images required for successful localization and the meaning of the vertical coordinate is the number of groups of tested sequences in this interval, when the number of images required for successful localization is greater than 60, we consider the localization has failed. From the figure, we can see that the image sequences enhanced by UMLE have a great improvement in both localization success and speed.}
\label{test2}
\end{figure}

\par In Fig.~\ref{fig:duibi}, the results demonstrate the importance of each of component with images, and the data in the table reflect the performance of each condition, where $w$ means without. Through the analysis of the image, we can see that the lack of a color discriminator and a texture discriminator  have a great impact on the color and texture. A multi-scale discriminator has a great influence on the whole image generation result. The complete model performs better in visual effect, and the results demonstrate the importance of each component.

\subsection{Localization  application}
Further, we used the proposed algorithm in a visual map localization application in scenarios with drastic changes in illumination (day$\rightarrow$night or light on$\rightarrow$off). In the experimental process, we use a publicly available SLAM dataset \cite{eth}, which is generated by the simulation environment and contains two sets of data for normal and flashlight illumination in the same scene.
\par The flowchart of our experiment is shown in Fig.~\ref{liuchengtu}. First, we randomly extract two types of images of normal light and flashlight from the dataset for training the UMLE network. At the same time, the sequences of continuous images under normal illumination are extracted to construct orb feature maps by using the ORB-SLAM2 \cite{ORBSLAM2}.
\par The experiments are mainly to validate our method for the localization problem under drastic light changes. In the localization process, for the pre-built feature map under normal light, we extract five sequence of flashlight images of the same scene at different locations, and try to use them for relocalization. Our localization method is implemented using the relocalization module in ORB-SLAM2 \cite{ORBSLAM2}. The results of feature point matching comparison before and after image enhancement can be seen in  Fig.~\ref{orbres}, and we can see that the enhanced image has a significant improvement on the ORB relocation results.
\par Fig.~\ref{test1} reflects the number of feature points before and after image enhancement. The number of images is the same, they all start from the same starting point at first. Since the number of successful feature points for many of the original image matches is 0, there is a steep drop at the beginning. We can see from the figure that the number of features matched in the enhanced image is significantly higher than in the original image.

\par We randomly selected 200 image sequences, each containing 60 images, from the flashlight dataset and the enhancement dataset to test the speed and success of relocalization of these images. The speed of image repositioning is reflected in Fig.~\ref{test2}. We consider a successful match within 5 frames as immediate relocation and a failed relocation above 60 frames.
\par From Fig.~\ref{test2}, we can see that the success rate of immediate re-localization of the image sequence enhanced with UMLE is 71.5\% compared to 47\% of the original image, and the success rate of re-localization of the enhanced image is much higher than that of the original image, and the failure rate of relocalization of the enhanced image is 0.15\% compared to 20\% of the original image, and the failure rate of re-localization of the enhanced image is much lower than that of the original image. This experiment verifies that the speed and success rate of re-localization of the enhanced image are significantly improved under our model.

\par As can be seen from the experimental results in the images, our proposed method is able to significantly improve the success rate of image matching for the same scene in the case of drastic lighting changes. Our method provides an effective solution to the visual localization problem in scenes with all-weather light changes and is of great practical value.

\subsection{Detection application}
%In the automatic driving task, illumination change has great influence on object detection, depth estimation and instance segmentation tasks, so we attempt to use our model  into the  conversion task between night scene and day scene. We test on the Oxford dataset \cite{30}. From the results which are shown in Fig.~\ref{fig:change}, we can see that our model can achieve a good conversion of lane line and building texture information. By the analysis of the performance  we can safely come to the conclusion that the proposed model could well transform the night scene into the day scene in the same scene, and demonstrate the feasible of our model in this issue.

\begin{figure}[t]
\begin{center}
\begin{tabular}{ccccc}
\hspace{-3.5mm}
\includegraphics[width = 0.49\linewidth]{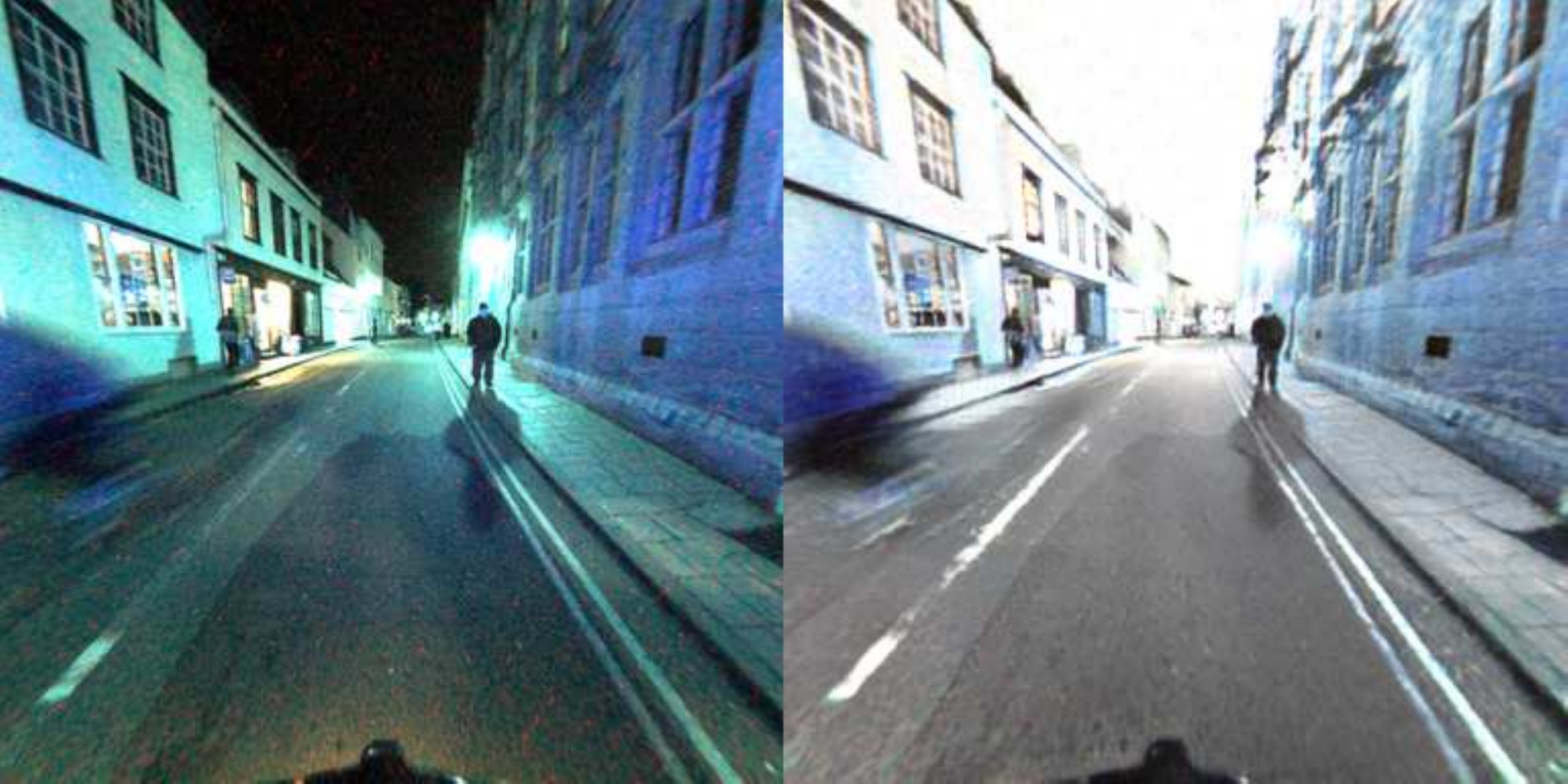} &\hspace{-5mm}
\includegraphics[width = 0.49\linewidth]{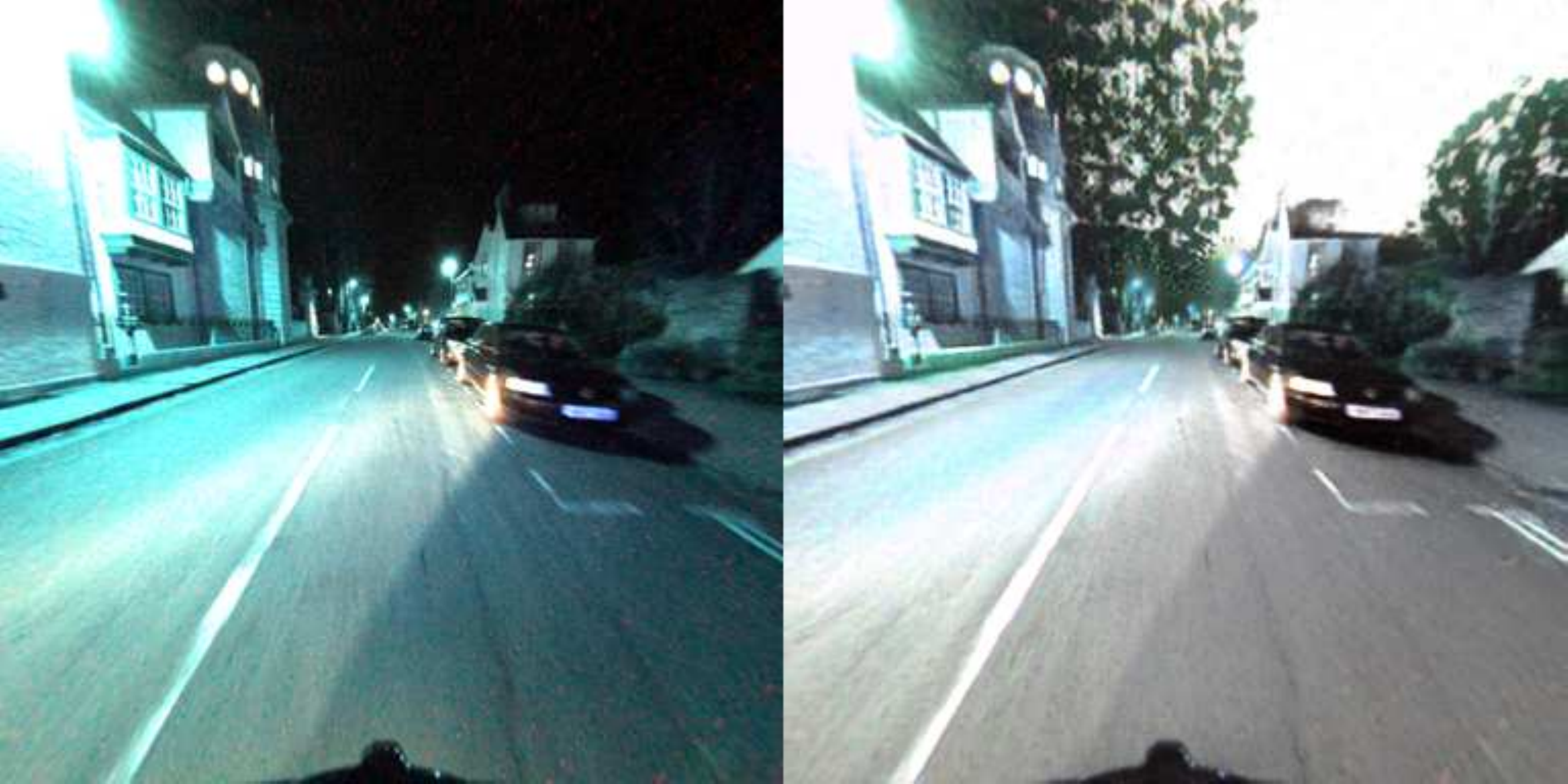} &\hspace{-5mm}
\\
\hspace{-3.5mm}
\includegraphics[width = 0.49\linewidth]{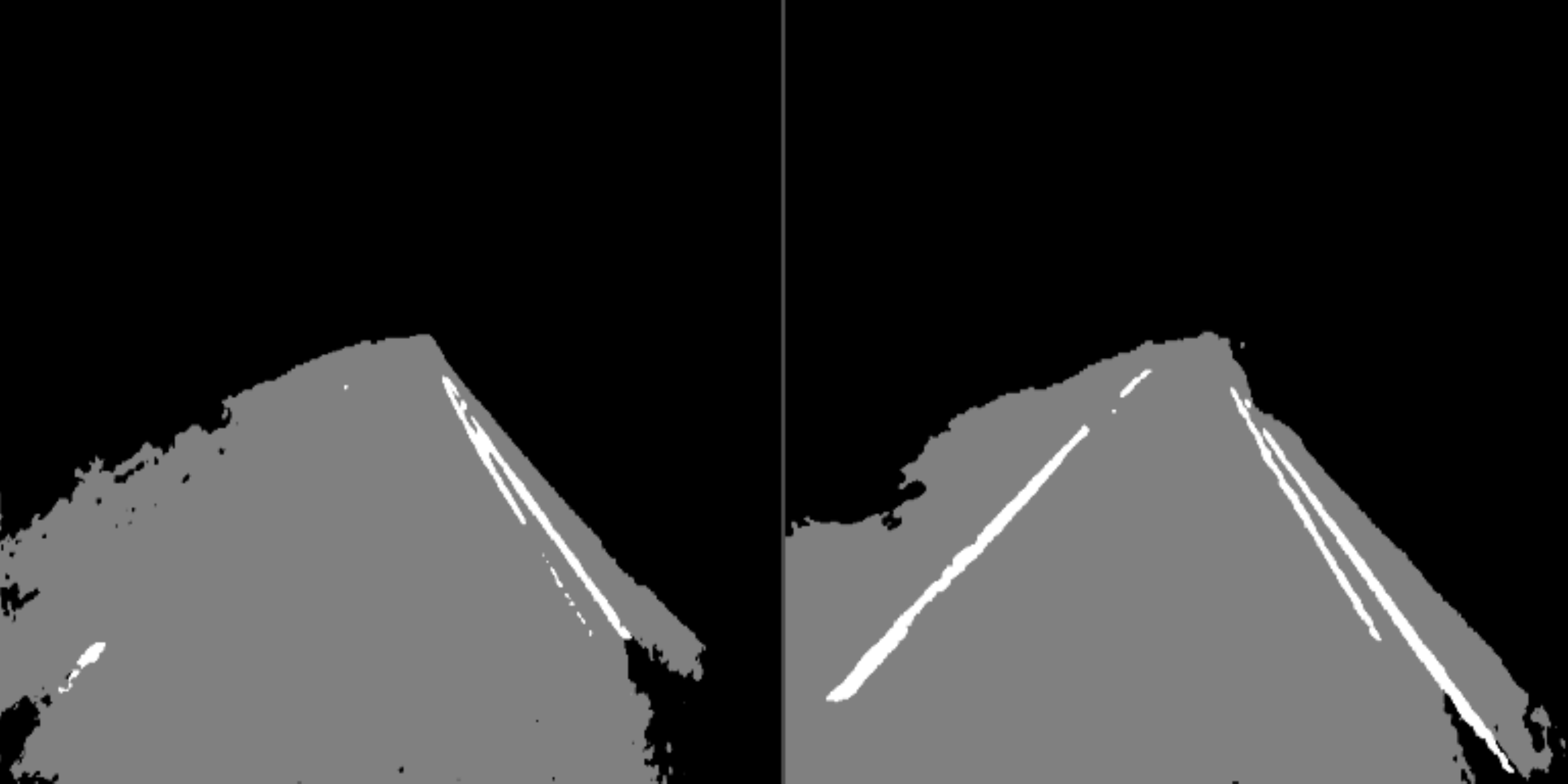} &\hspace{-5mm}
\includegraphics[width = 0.49\linewidth]{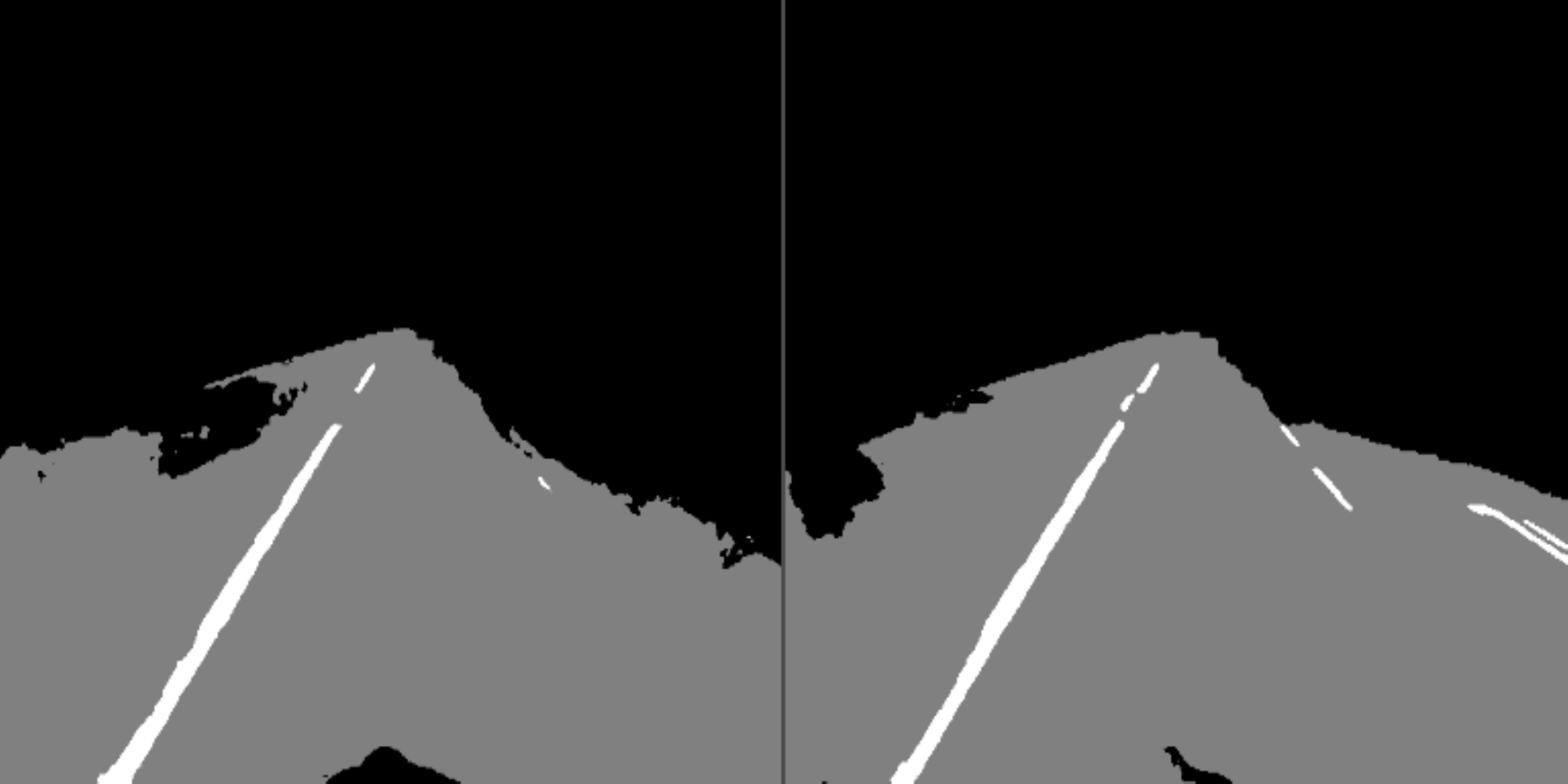} &\hspace{-5mm}
\\
(a) &\hspace{-4mm} (b)
\\
\end{tabular}
\end{center}
\caption{ Driveable area detection results before and after image enhancement. Images in the first row are the night image and the corresponding results enhanced by UMLE; the second row are the detection results of the drivable area of the corresponding images in the first row. }
\label{fig:change}
\end{figure}

Illumination changes have a significant impact on driveable area detection in the autopilot task, so we attempt to apply our model to a night and day conversion task. We test on the Oxford dataset \cite{30}. From the results which are shown in Fig.~\ref{fig:change}, it can be seen that UMLE can achieve a good conversion of image information and a more realistic detection of drivable areas as well as lane lines. By analyzing the performance, we can safely conclude that the proposed model can convert night scenes into day scenes in the same scene and improve the driveable area detection results very well.

\section{Conclusion}
In this paper, we  propose an unsupervised enhancement model for real-time low-light image enhancement. The model's training is independent of paired  training data, so it can use images of different scenes and different illuminations. Besides, we present a multi-branch discriminator which can comprehensively evaluate the image  from color, texture, and global information.  We also introduce a novel attention module which combines the channel attention  and the pixel attention to help to focus more on the low-light areas. Both qualitative and quantitative experimental evaluations show that our network achieves good results in terms of visual effects and noise control. In addition, we experimentally validate that our method is significantly more effective for tasks such as SLAM repositioning drivable area detection.

\bigskip
\bibliographystyle{IEEEtran}
\bibliography{egbib}

% Generated by IEEEtran.bst, version: 1.13 (2008/09/30)
\begin{thebibliography}{10}
\providecommand{\url}[1]{#1}
\csname url@samestyle\endcsname
\providecommand{\newblock}{\relax}
\providecommand{\bibinfo}[2]{#2}
\providecommand{\BIBentrySTDinterwordspacing}{\spaceskip=0pt\relax}
\providecommand{\BIBentryALTinterwordstretchfactor}{4}
\providecommand{\BIBentryALTinterwordspacing}{\spaceskip=\fontdimen2\font plus
\BIBentryALTinterwordstretchfactor\fontdimen3\font minus
  \fontdimen4\font\relax}
\providecommand{\BIBforeignlanguage}[2]{{%
\expandafter\ifx\csname l@#1\endcsname\relax
\typeout{** WARNING: IEEEtran.bst: No hyphenation pattern has been}%
\typeout{** loaded for the language `#1'. Using the pattern for}%
\typeout{** the default language instead.}%
\else
\language=\csname l@#1\endcsname
\fi
#2}}
\providecommand{\BIBdecl}{\relax}
\BIBdecl

\bibitem{2}
S.~M. Pizer, E.~P. Amburn, J.~D. Austin, R.~Cromartie, and K.~Zuiderveld,
  ``Adaptive histogram equalization and its variations,'' \emph{Computer Vision
  Graphics \& Image Processing}, vol.~39, no.~3, pp. 355--368, 1987.

\bibitem{5}
H.~Ibrahim and N.~S.~P. Kong, ``Brightness preserving dynamic histogram
  equalization for image contrast enhancement,'' \emph{IEEE Transactions on
  Consumer Electronics}, vol.~53, no.~4, pp. 1752--1758, 2008.

\bibitem{9}
C.~Lee, C.~Lee, and C.~Kim, ``Contrast enhancement based on layered difference
  representation of 2d histograms,'' \emph{{IEEE} Trans. Image Process.},
  vol.~22, no.~12, pp. 5372--5384, 2013.

\bibitem{3}
E.~H. Land, ``The retinex theory of color vision,'' \emph{Scientific american},
  vol. 237, no.~6, pp. 108--129, 1977.

\bibitem{11}
D.~J. Jobson, Z.~Rahman, and G.~A. Woodell, ``A multiscale retinex for bridging
  the gap between color images and the human observation of scenes,''
  \emph{IEEE Transactions on Image processing}, vol.~6, no.~7, pp. 965--976,
  1997.

\bibitem{10}
X.~Guo, ``{LIME:} {A} method for low-light image enhancement,'' in
  \emph{Proceedings of the 2016 {ACM} Conference on Multimedia Conference},
  A.~Hanjalic, C.~Snoek, M.~Worring, D.~C.~A. Bulterman, B.~Huet, A.~Kelliher,
  Y.~Kompatsiaris, and J.~Li, Eds., 2016, pp. 87--91.

\bibitem{7}
K.~Nakai, Y.~Hoshi, and A.~Taguchi, ``Color image contrast enhacement method
  based on differential intensity/saturation gray-levels histograms,'' in
  \emph{International Symposium on Intelligent Signal Processing and
  Communication Systems}, 2013, pp. 445--449.

\bibitem{17}
Z.~Ying, G.~Li, and W.~Gao, ``A bio-inspired multi-exposure fusion framework
  for low-light image enhancement,'' \emph{CoRR}, vol. abs/1711.00591, 2017.

\bibitem{24}
K.~G. Lore, A.~Akintayo, and S.~Sarkar, ``Llnet: A deep autoencoder approach to
  natural low-light image enhancement,'' \emph{Pattern Recognition}, vol.~61,
  pp. 650--662, 2015.

\bibitem{62}
C.~Wei, W.~Wang, W.~Yang, and J.~Liu, ``Deep retinex decomposition for
  low-light enhancement,'' \emph{arXiv preprint arXiv:1808.04560}, 2018.

\bibitem{26}
C.~Chen, Q.~Chen, J.~Xu, and V.~Koltun, ``Learning to see in the dark,'' in
  \emph{Proceedings of the IEEE Conference on Computer Vision and Pattern
  Recognition}, 2018, pp. 3291--3300.

\bibitem{27}
Y.~Zhang, J.~Zhang, and X.~Guo, ``Kindling the darkness: A practical low-light
  image enhancer,'' in \emph{Proceedings of the 27th ACM International
  Conference on Multimedia}, 2019, pp. 1632--1640.

\bibitem{qu}
Y.~Qu, Y.~Ou, and R.~Xiong, ``Low illumination enhancement for object detection
  in self-driving,'' in \emph{2019 IEEE International Conference on Robotics
  and Biomimetics (ROBIO)}.\hskip 1em plus 0.5em minus 0.4em\relax IEEE, 2019,
  pp. 1738--1743.

\bibitem{NICE}
R.~Chen, W.~Huang, B.~Huang, F.~Sun, and B.~Fang, ``Reusing discriminators for
  encoding: Towards unsupervised image-to-image translation,'' in \emph{2020
  {IEEE/CVF} Conference on Computer Vision and Pattern Recognition}, 2020, pp.
  8165--8174.

\bibitem{SENet}
J.~Hu, L.~Shen, S.~Albanie, G.~Sun, and E.~Wu, ``Squeeze-and-excitation
  networks,'' \emph{{IEEE} Trans. Pattern Anal. Mach. Intell.}, vol.~42, no.~8,
  pp. 2011--2023, 2020.

\bibitem{ECA-Net}
Q.~Wang, B.~Wu, P.~Zhu, P.~Li, W.~Zuo, and Q.~Hu, ``Eca-net: Efficient channel
  attention for deep convolutional neural networks,'' in \emph{2020 {IEEE/CVF}
  Conference on Computer Vision and Pattern Recognition, {CVPR}}.\hskip 1em
  plus 0.5em minus 0.4em\relax {IEEE}, 2020, pp. 11\,531--11\,539.

\bibitem{30}
A.~Anoosheh, T.~Sattler, R.~Timofte, M.~Pollefeys, and L.~Van~Gool,
  ``Night-to-day image translation for retrieval-based localization,'' in
  \emph{2019 International Conference on Robotics and Automation (ICRA)}.\hskip
  1em plus 0.5em minus 0.4em\relax IEEE, 2019, pp. 5958--5964.

\bibitem{31}
Y.~Jiang, X.~Gong, D.~Liu, Y.~Cheng, C.~Fang, X.~Shen, J.~Yang, P.~Zhou, and
  Z.~Wang, ``Enlightengan: Deep light enhancement without paired supervision,''
  \emph{CoRR}, vol. abs/1906.06972, 2019.

\bibitem{lsgan}
X.~Mao, Q.~Li, H.~Xie, R.~Y.~K. Lau, Z.~Wang, and S.~P. Smolley, ``Least
  squares generative adversarial networks,'' in \emph{{IEEE} International
  Conference on Computer Vision, {ICCV}}.\hskip 1em plus 0.5em minus
  0.4em\relax {IEEE} Computer Society, 2017, pp. 2813--2821.

\bibitem{dataset2}
N.~K. Kalantari and R.~Ramamoorthi, ``Deep high dynamic range imaging of
  dynamic scenes,'' \emph{{ACM} Trans. Graph.}, vol.~36, no.~4, pp.
  144:1--144:12, 2017.

\bibitem{relu}
X.~Glorot, A.~Bordes, and Y.~Bengio, ``Deep sparse rectifier neural networks,''
  in \emph{Proceedings of the Fourteenth International Conference on Artificial
  Intelligence and Statistics}, vol.~15, 2011, pp. 315--323.

\bibitem{AdamGC}
H.~Yong, J.~Huang, X.~Hua, and L.~Zhang, ``Gradient centralization: {A} new
  optimization technique for deep neural networks,'' \emph{CoRR}, vol.
  abs/2004.01461, 2020.

\bibitem{49}
J.~Kim, M.~Kim, H.~Kang, and K.~Lee, ``U-gat-it: unsupervised generative
  attentional networks with adaptive layer-instance normalization for
  image-to-image translation,'' \emph{arXiv preprint arXiv:1907.10830}, 2019.

\bibitem{54}
X.~Huang and S.~J. Belongie, ``Arbitrary style transfer in real-time with
  adaptive instance normalization,'' in \emph{{IEEE} International Conference
  on Computer Vision, {ICCV}}.\hskip 1em plus 0.5em minus 0.4em\relax {IEEE}
  Computer Society, 2017, pp. 1510--1519.

\bibitem{53}
L.~J. Ba, J.~R. Kiros, and G.~E. Hinton, ``Layer normalization,'' \emph{CoRR},
  vol. abs/1607.06450, 2016.

\bibitem{MBLLEN}
F.~Lv, F.~Lu, J.~Wu, and C.~Lim, ``{MBLLEN:} low-light image/video enhancement
  using cnns,'' in \emph{British Machine Vision Conference 2018}, 2018, p. 220.

\bibitem{46}
J.-Y. Zhu, T.~Park, P.~Isola, and A.~A. Efros, ``Unpaired image-to-image
  translation using cycle-consistent adversarial networks,'' in
  \emph{Proceedings of the IEEE international conference on computer vision},
  2017, pp. 2223--2232.

\bibitem{GLAD}
W.~Wang, C.~Wei, W.~Yang, and J.~Liu, ``Gladnet: Low-light enhancement network
  with global awareness,'' in \emph{13th {IEEE} International Conference on
  Automatic Face {\&} Gesture}, 2018, pp. 751--755.

\bibitem{NIQE}
A.~Mittal, Fellow, IEEE, R.~Soundararajan, and A.~C. Bovik, ``Making a
  'completely blind' image quality analyzer,'' \emph{IEEE Signal Processing
  Letters}, vol.~20, no.~3, pp. 209--212, 2013.

\bibitem{entropy}
J.~R. A, ``Focus optimization criteria for computer image processing [j],'' in
  \emph{Microscope}, 1976, pp. 163--180.

\bibitem{eth}
S.~Park, T.~Sch\"ops, and M.~Pollefeys, ``Illumination change robustness in
  direct visual slam,'' in \emph{ICRA}, 2017.

\bibitem{ORBSLAM2}
R.~Mur~Artal and J.~D. Tard\'os, ``{ORB-SLAM2}: an open-source {SLAM} system
  for monocular, stereo and {RGB-D} cameras,'' \emph{IEEE Transactions on
  Robotics}, vol.~33, no.~5, pp. 1255--1262, 2017.

\bibitem{yolo}
\BIBentryALTinterwordspacing
J.~Redmon and A.~Farhadi, ``Yolov3: An incremental improvement,'' \emph{CoRR},
  vol. abs/1804.02767, 2018. [Online]. Available:
  \url{http://arxiv.org/abs/1804.02767}
\BIBentrySTDinterwordspacing

\end{thebibliography}

\end{document}